\documentclass[preprint]{aastex}

\usepackage{natbib}
\usepackage{graphicx}
\usepackage{rotating}
\usepackage{enumitem}
\usepackage{amssymb,amsmath}

\newcommand{\Rs}{$ R_{\odot}$}

\newcommand{\de}{$^{\circ}$}

\newcommand{\pB}{\textit{pB}}

\newcommand{\kms}{~km$s^{-1}$}
\newcommand{\mss}{~m$s^{-2}$}

\newcommand{\cmvol}{cm$^{-3}$}
\newcommand{\paperi}{Paper I}
\newcommand{\paperii}{Paper II}
\makeatletter

\newcommand{\Rmnum}[1]{\expandafter\@slowromancap\romannumeral #1@}

\makeatother

\begin{document}

\shorttitle{Characteristics of coronal streamers}
\shortauthors{Morgan and Cook}
\title{The width, density and outflow of solar coronal streamers.}
\author{Huw Morgan} 
\author{Anthony C. Cook}
\affil{Institute of Mathematics, Physics and Computer Science, Aberystwyth University, Ceredigion, Cymru, SY23 3BZ}
\email{hmorgan@aber.ac.uk}

\begin{abstract}
Characterising the large-scale structure and plasma properties of the inner corona is crucial to understand the source and subsequent expansion of the solar wind and related space weather effects. Here we apply a new coronal rotational tomography method, along with a method to narrow streamers and refine the density estimate, to COR2A/STEREO observations from a period near solar minimum and maximum, gaining density maps for heights between 4 and 8\Rs. The coronal structure is highly radial at these heights, and the streamers are very narrow, in some regions only a few degrees in width. The mean densities of streamers is almost identical between solar minimum and maximum. However, streamers at solar maximum contain around 50\%\ more total mass due to their larger area. By assuming a constant mass flux, and constraints on proton flux measured by Parker Solar Probe (PSP), we estimate an outflow speed within solar minimum streamers of 50-120\kms\ at 4\Rs, increasing to 90-250\kms\ at 8\Rs. Accelerations of around 6\mss\ are found for streamers at a height of 4\Rs, decreasing with height. The solar maximum slow wind shows a higher acceleration to extended distances compared to solar minimum.  To satisfy the solar wind speeds measured by PSP,  there must be a mean residual acceleration of around 1-2\mss\ between 8 and 40\Rs. Several aspects of this study strongly suggest that the coronal streamer belt density is highly variable on small scales, and that the tomography can only reveal a local spatial and temporal average.
\end{abstract}

\maketitle

\section{Introduction}

Coronal rotational tomography (CRT) of coronagraph imaging data currently offers the only direct method to reconstruct the global coronal density structure. CRT provides unique insight into: (i) how the density and structure vary with height from the low corona observed in Extreme UltraViolet \citep[e.g.][]{vasquez2016}, to the extended inner corona \citep[e.g.][]{morgan2019, frazin2010}, and regions in between \citep[e.g.][]{kramar2009}; (ii) the short-term \citep{morgan2011solmin,vibert2016} and long-term \citep{morgan2010structure, morgan2011rotation} time evolution of the corona; and (iii) the relationship between magnetic and plasma structures \citep{vasquez2008,morgan2010structure,morgan2011solmin,kramar2014}. Furthermore, estimates of the 3D distribution of density is crucial for interpretation of observations by other instruments \citep{frazin2003}, and as boundary constraints for large-scale models \citep{frazin2005a}. These methods, and their findings, are described in the review of \citet{aschwanden2011}. Most of the literature is dedicated to descriptions and tests of methods, with application to a number of case studies. Exceptions that look at larger datasets and longer periods are \citet{morgan2010structure} and \citet{morgan2011rotation}, where more long-term studies of the streamer distribution are made, albeit based on a qualitative view of density structure. CRT of the solar maximum corona is particularly difficult, and results from these periods are rare \citep{butala2005,morgan2010structure}. 

Recently, \citet{morgan2019} (hereafter \paperii) developed a new CRT method based on spherical harmonics. The success of the method depends on the detailed calibration and pre-processing methods described by \citet{morgan2015} (hereafter \paperi). The method is restricted to heights where the coronal structure is predominantly radial, so at heliocentric distances of 3\Rs\ or greater. In this work, section \ref{method} describes the application of the new method to observations. An extension and refinement to the CRT method is also presented in section \ref{method}, necessary to replicate the apparent narrowness of streamers in the coronagraph images. Results for both solar minimum and maximum are presented in section \ref{results} for a range of heights between 4 and 8\Rs. These results lead to a comparison of streamer and coronal hole mean densities as functions of height at solar minimum and maximum in section \ref{sectiondensity}, and a comparison to previously published values. A consideration of mass flux and constraints on outflow speeds and acceleration are given in section \ref{sectionmass}.  Section \ref{sectionmap} presents a new mapping which visualises the 3D density alongside the coronagraph images. A summary of results, and overview of future work, is given in section \ref{conclusions}. 

\section{Tomography maps and streamer narrowing method}
\label{method}

\begin{figure}[]
\begin{center}
\includegraphics[width=8.5cm]{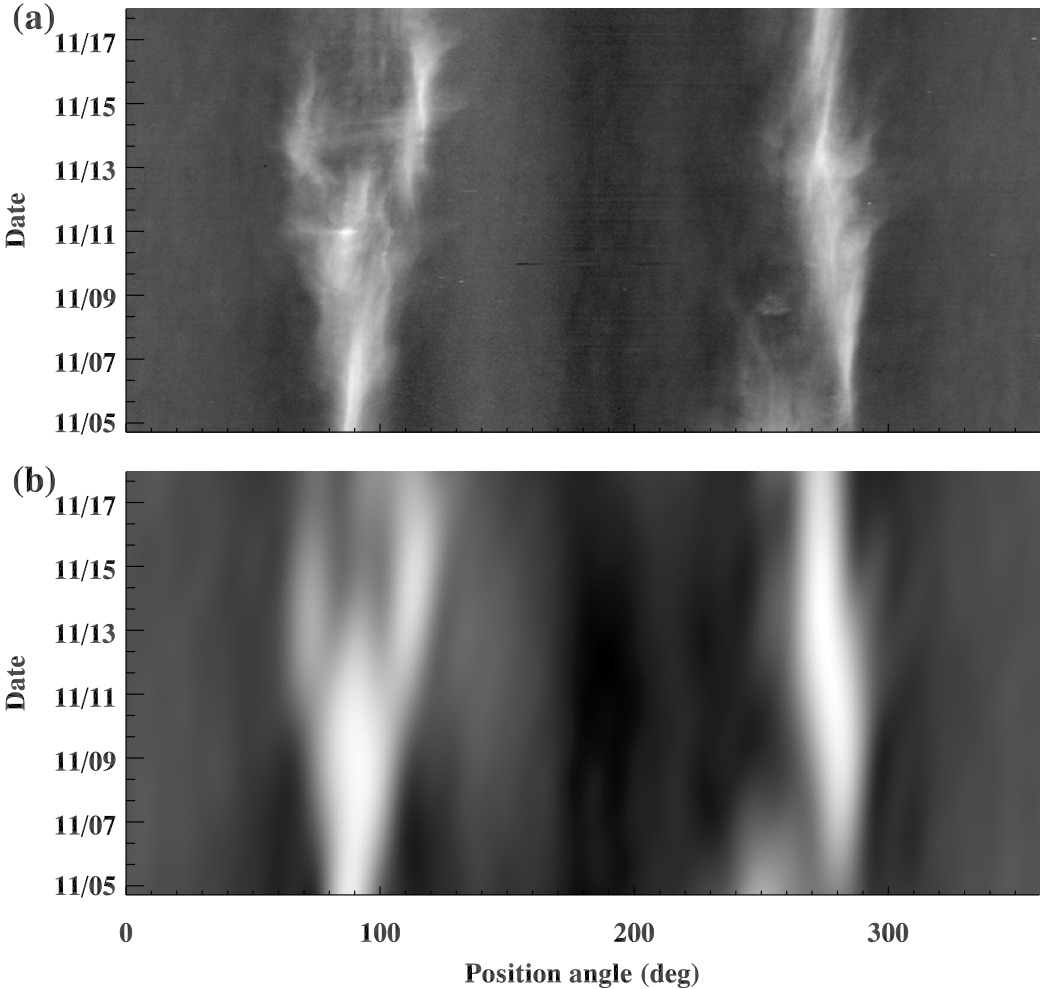}
\end{center}
\caption{Polarized brightness as a function of position angle and observation time at heights of 6\Rs\ for (a) COR2A observations and (b) as gained from the tomography density map.}
\label{bmethod}
\end{figure}

The COR2 coronagraphs are part of the Sun Earth Connection Coronal and Heliospheric Investigation (SECCHI, \citet{howard2002}) suite of instruments aboard the twin Solar Terrestial Relations Observatory (STEREO A \& B, \citet{kaiser2005}). Initial density maps are created from COR2A data using the calibration procedures of \paperi\ and the spherical-harmonic tomography method of \paperii. The COR2 polarized brightness (\pB) observations over a period of half a Carrington rotation ($\pm$1 week from the required mid-date) are used, conveniently saved locally in polar co-ordinate format. The tomography takes as input circular slices of \pB\ data at a constant heliocentric height over the $\approx$2-week period. An example of input data for mid-date 2018/11/11 and height 6\Rs\ is shown in figure \ref{bmethod}a. The tomography method of \paperii\ is applied with a $25^{th}$ order spherical harmonic basis to create density reconstructions with 540 longitude and 270 latitude bins for a set of 9 heights between 4 and 8\Rs, at 0.5\Rs\ increments. The high resolution of input data, and high order of spherical harmonics, is applied using high-performance computing, provided by the facilities and support of Super Computing Wales (see acknowledgments). 

\begin{figure}[]
\begin{center}
\includegraphics[width=8.5cm]{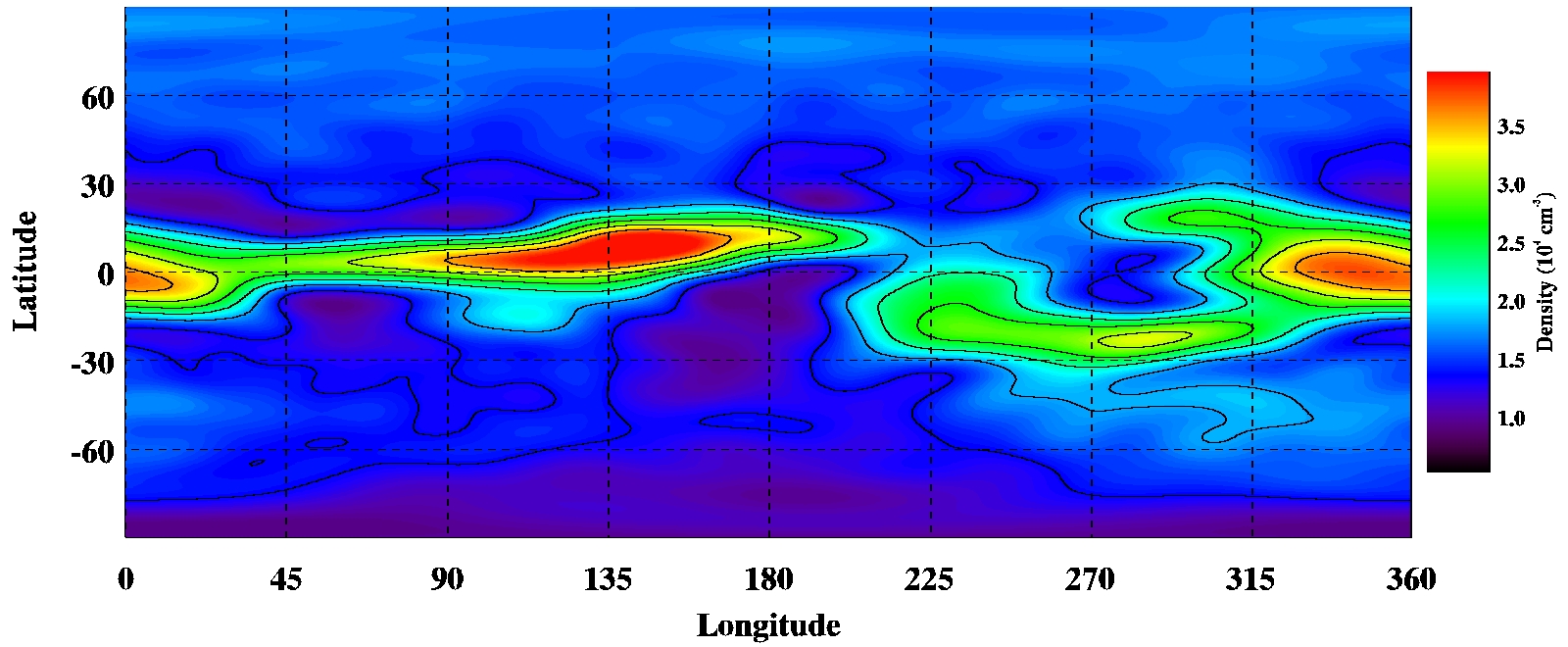}
\end{center}
\caption{Density map of the solar corona at height 6.0\Rs\ for mid-date 2018/11/11 in Carrington longitude-latitude coordinates formed using the spherical harmonic approach of \paperii. The colours in the map correspond to the densities shown in the color bar, in units of $10^{-4}$\cmvol.}
\label{densmethod}
\end{figure}

An example of a density map, for mid-date 2018/11/11 and heliocentric distance 6.0\Rs, is shown in figure \ref{densmethod}. The streamer belt is restricted to the equator, and meanders with only a gradual latitudinal variation. For longitudes between 180 and 315\de, the streamer belt is split into two branches, as is commonly observed at times close to solar minimum \citep{morgan2011solmin}. The south branch is the main current sheet or neutral line streamer, the north is a pseudostreamer. Densities outside of the streamers is around $0.9-1.8 \times 10^4$ \cmvol, with peak streamer densities at $3.6 \times 10^4$ \cmvol.

The observed brightness for this period and height is shown in figure \ref{bmethod}a. The reconstructed brightness as given by the tomography density map is shown in figure \ref{bmethod}b. Whilst the position of the streamer belt, the main splitting of the belt, and other fainter structures, are well replicated by the tomography, the reconstructed brightness has less fine-scale structure and the streamer belt is generally far too wide in position angle. It is apparent therefore that one general defect of the tomography is that the reconstructed streamers are too wide, or that the density structure is overly smooth. This is an unavoidable effect of the tomography method, where smoothness of the reconstruction is a required criteria (see \paperii). 

In order to investigate the true width of streamers, features in the tomography maps are `sharpened' or narrowed using a two-step process. The first step uses the gradients in the density maps to sharpen high-density features, and the second compares observed with model brightness to improve the fit.

\subsection{Sharpening the density structure}

The longitude-latitude density map is extended in longitude to avoid edge effects at the 0/360\de\ boundary, so that longitudes extend from $-10$ to 370\de, with a repeat of density structure at both longitudinal edges. Longitudinal and latitudinal normalized density gradients are calculated as the first forward difference in longitude and latitude respectively:
\begin{align}
g_{i,j} &= \rho^\prime_{i,j} - \rho^\prime_{i-1,j} \\
h_{i,j} &= \rho^\prime_{i,j} - \rho^\prime_{i,j-1} \,, 
\end{align}
where $\rho^\prime$ is the density normalized to values between zero and unity according to the density minimum and maximum value, and the $i,j$ subscripts give the longitudinal and latitudinal bin index. Indices $i^\prime$ and $j^\prime$, at which density values will be interpolated from the original density map, are defined as
\begin{align}
i^\prime &= i - fg_{i,j} \\
j^\prime &= j - fh_{i,j} ,
\end{align}
where $f$ is a value that controls the sharpening of density features in the output. The output density at bin $i,j$ is interpolated from the original densities at positions $i^\prime, j^\prime$ using cubic bilinear interpolation. Values of $f$ range from 0 (no change between original and output density map) and 300 (streamers become very narrow). The densities in the output map are multiplied by a factor in order to preserve the total integrated density of the original map.

\begin{figure}[]
\begin{center}
\includegraphics[width=8.5cm]{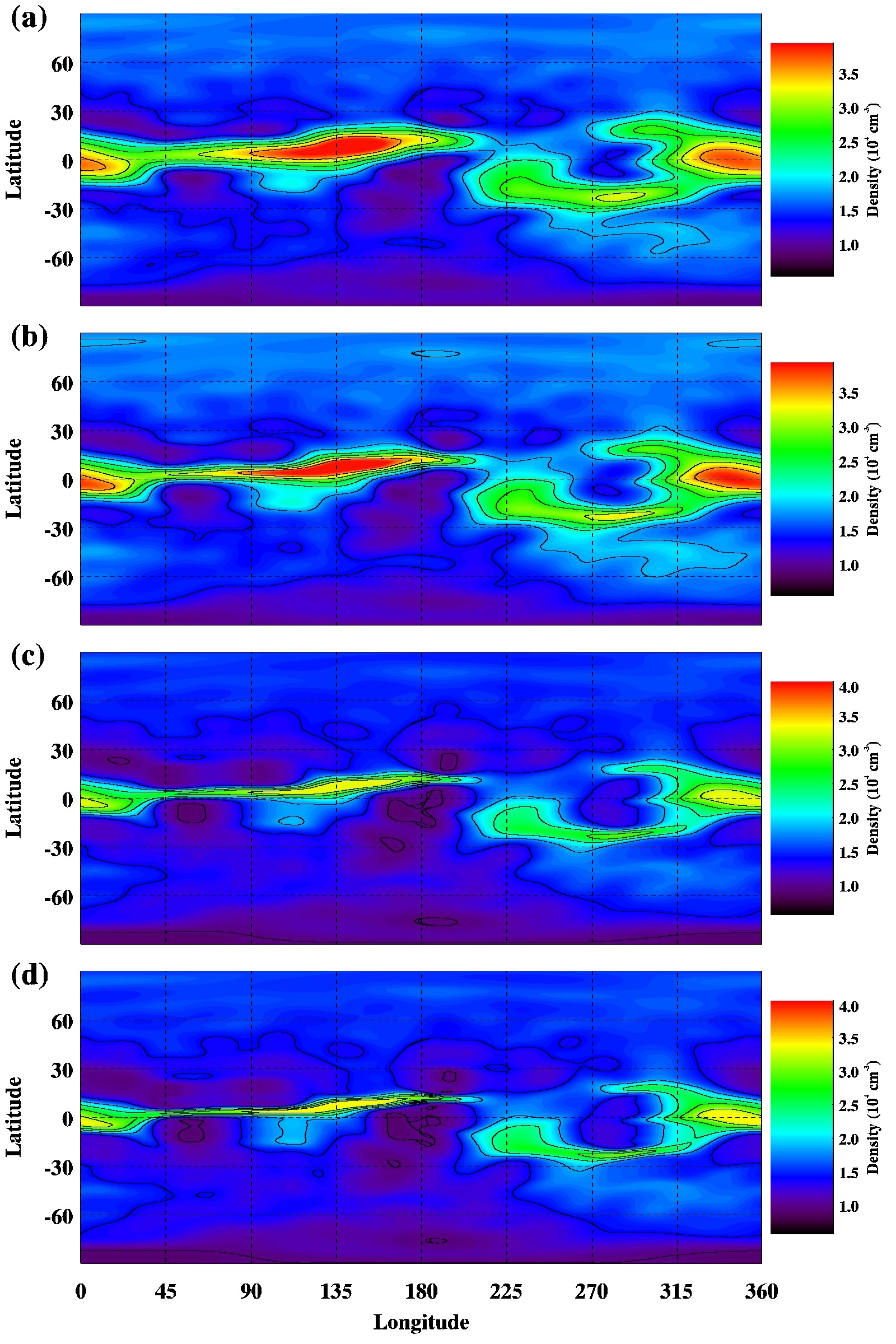}
\end{center}
\caption{Density maps gained by applying the sharpening method for (a)-(d) $f=0,100,200,300$ respectively to the original density map of figure \ref{densmethod}. Note that (a) $f=0$ is identical to the original map.}
\label{densnarrow}
\end{figure}

Examples of the densities gained after applying the sharpening process for various values of $f$ are shown in figure \ref{densnarrow}a-d. The method is effective at narrowing the streamer structures without introducing artifacts. The maximum densities increase as the streamers become narrower due to preserving the same integrated densities across the whole map. Note that this method works on the smooth, large-scale structures of the density map, but would face problems on images with higher-frequency components.

The act of narrowing the density structures degrades the optimal fit found between reconstructed and observed brightness by the tomography, thus a further process is applied that iteratively optimises the narrowed densities to improve the fit. The process takes local regions of the density map, map these regions to areas of the reconstructed brightness using line-of-sight (LOS) integrations, calculates the ratio of reconstructed and observed brightness, and maps this ratio back to the local density region. Repeated over many regions and combined, a correction is applied to the density. This process is then repeated until convergence is reached when the change in density over iterations becomes small. Adjustment of streamer and coronal hole densities are separated in the process, so that the adjustment of coronal hole densities is not strongly influenced by the brighter streamer regions. Details of this iterative process is described in appendix \ref{app1}. This refinement method is applied to eleven different values of the streamer narrowing factor $f$, allowing identification of an optimal factor, described in appendix \ref{app2}.

\subsection{Correcting for excess density}
\label{fcoronasec}
The F-corona is unpolarized at low coronal heights, but has a polarized component which can become significant above $\approx$6\Rs\ \citep{mann1996}. This polarized component is poorly understood, and lacks estimates based on measurements at heights of interest in the extended inner corona. For this reason, no correction has been made to the coronagraph data. However, the tomography density maps give a way to estimate a lower limit to the F-coronal polarized brightness, and/or other possible calibration errors, based on simple physical arguments on the evolution of the plasma density with height.

\begin{figure}[]
\begin{center}
\includegraphics[width=7.5cm]{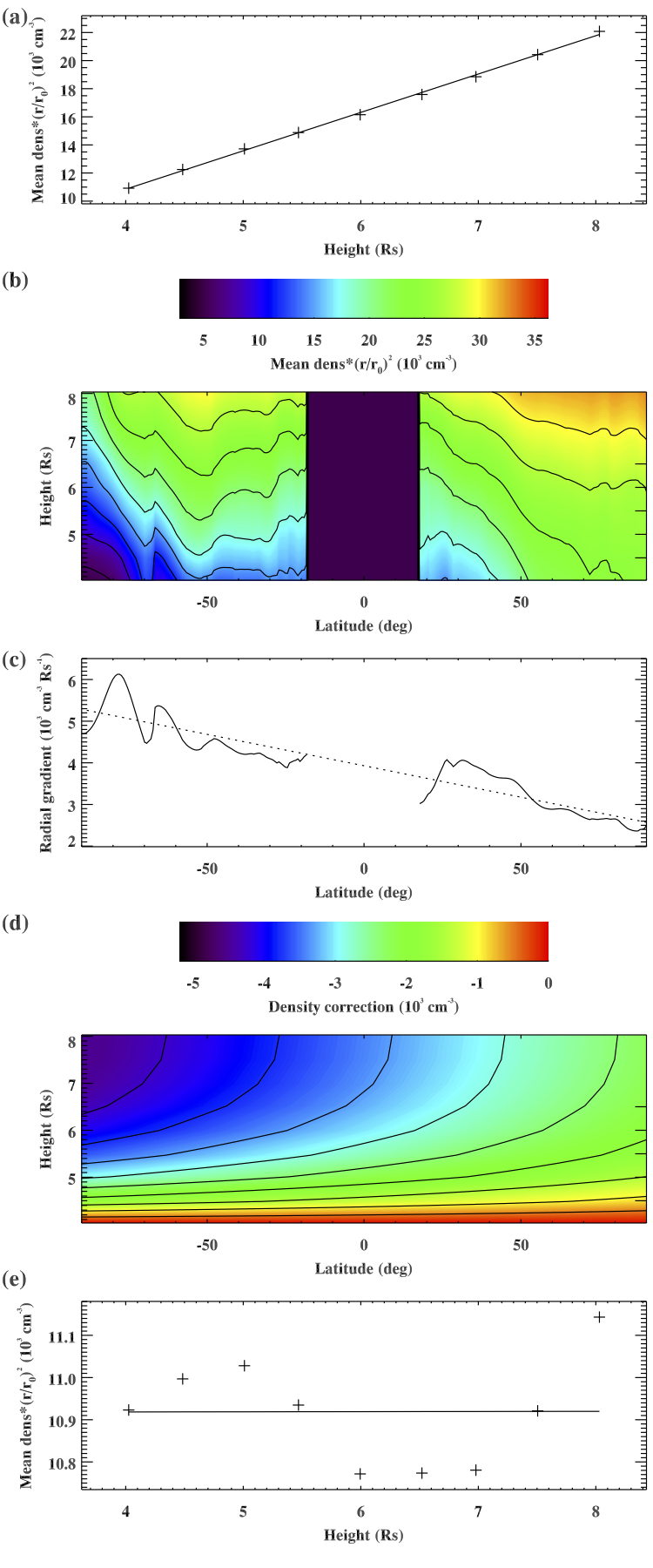}
\end{center}
\caption{(a) The value $\varrho = \bar{\rho}_{ch} (r/r_0)^2$ calculated for tomography density maps at several heights (see text). (b) $\varrho$ calculated as a function of latitude and height. The missing data near the equator is due to the low area of coronal holes, which gives unreliable results. (c) The radial gradient of $\varrho$ as a function of latitude (solid line), gained from fitting $\varrho$ to a $1^{st}$ order polynomial of height. The dotted line shows a linear fit of this gradient to latitude. (d) The density correction that is applied to the tomography maps, leading to a more constant $\varrho$ over latitude and height. (e) $\varrho$ as a function of height following the density correction.}
\label{fixdens1}
\end{figure}

The tomography, streamer-narrowing, and density refinement method is applied to the 2018/11/11 dataset resulting in a set of tomography maps for heights between 4 and 8\Rs\ in 0.5\Rs\ increments. At a given height, the mean density within coronal holes, $\bar{\rho}_{ch}$, is calculated. Assuming a radial expansion of the corona, under constant outflow speed, and with a constant mass flux, the value $\varrho = \bar{\rho}_{ch} (r/r_0)^2$ should be constant at all heights, from the base height $r_0=4$\Rs\ outwards. If an accelerating expansion is assumed, $\varrho$ should decrease with height. The value $\varrho$ is shown in figure \ref{fixdens1}a. There is a $\approx$50\%\ linear increase of $\varrho$ between 4 and 8\Rs. The most likely cause of this increase is the increasing polarization of the F-corona with height. 

Figure \ref{fixdens1}b shows $\varrho$ as a function of latitude and height. Except for equatorial streamer regions, where the results are unreliable and have been omitted, $\varrho$ shows a close to linear increase at all heights. This positive gradient is shown as a function of latitude in figure \ref{fixdens1}c, gained from fitting the height profiles of $\varrho$ at each latitude to a $1^{st}$ order polynomial. The gradient is positive at all latitudes, and there is an approximately linear decrease in the gradient from the south to the north. This allows a linear fitting of the gradient to latitude, shown as the dotted line. This gradient leads to a density correction, shown in figure \ref{fixdens1}d, that can be applied to the tomography maps, and will lead to a more constant $\varrho$ with height. Figure \ref{fixdens1}e shows $\varrho$ after the correction. The value at the base height of 4\Rs\ is unchanged from the original density, and $\varrho$ remains almost constant with height.

The two most likely contributions to the excess density with increasing height are the polarized component of the F-corona, and an instrument calibration error (flat field error). There is no doubt that some component of the excess density is due to the polarized component of the F-corona since no attempt is made, prior to the density inversion, to correct for this. The different gradient of $\varrho$ with height is close to a linear relationship with latitude, seen in figure \ref{fixdens1}b and c. This leads to a density correction which is also dependent on latitude (greatest correction in the south, least in the north). The F-corona polarized brightness does not have a north-south difference (see \citet{morgan2007fcorona}), thus this gradient is solely due to the calibration error. 

\begin{figure}[]
\begin{center}
\includegraphics[width=7.5cm]{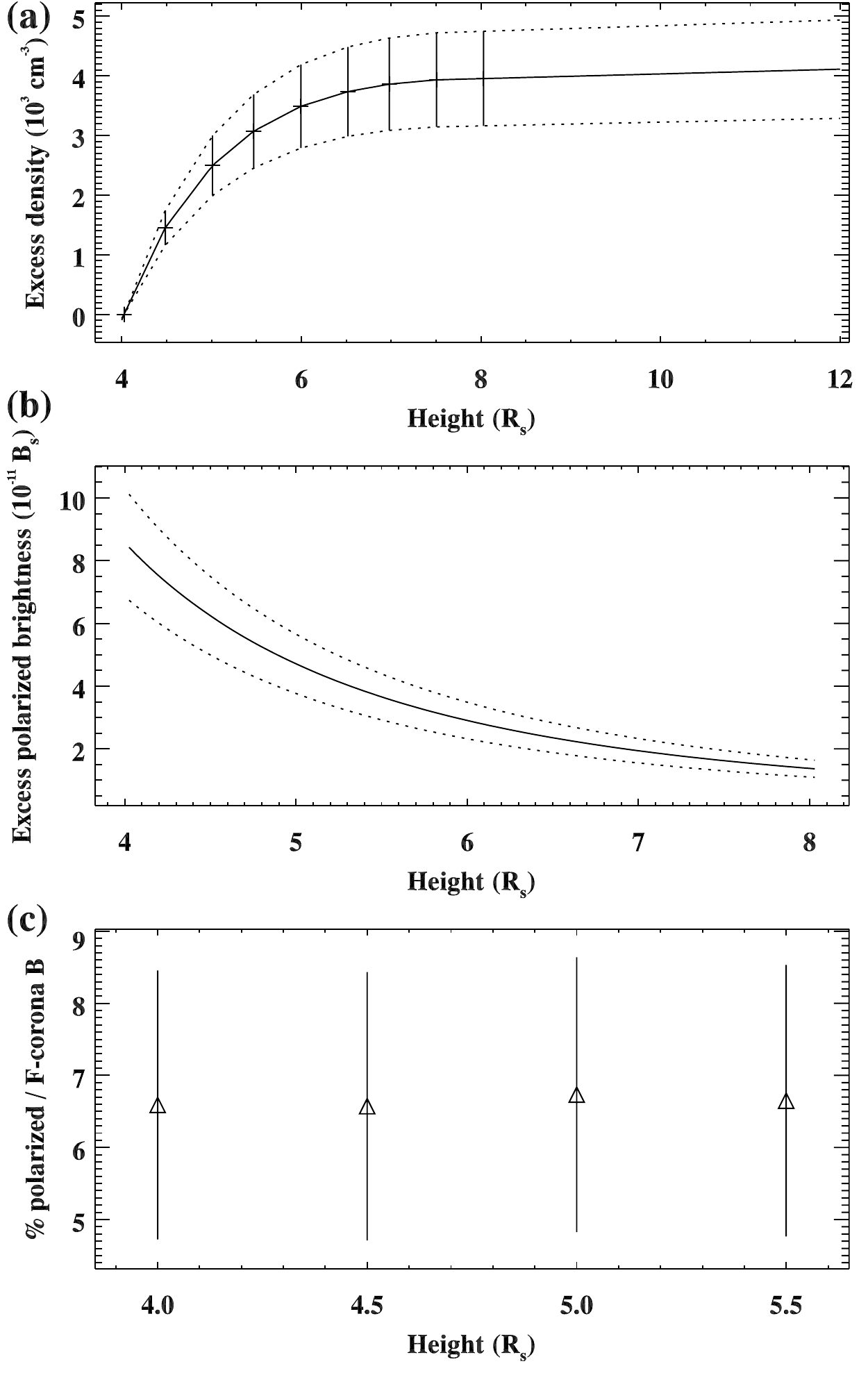}
\end{center}
\caption{(a) The mean excess density within the tomography maps (crosses), with error bars showing the variance over latitude. The solid line shows an interpolation between the data points and extrapolation out to 12\Rs. The dotted lines show the extrapolated uncertainty. (b) The polarized brightness expected from a spherically-symmetric distribution of the excess density shown in (a). (c) The percentage ratio of polarized to the total F-coronal brightness estimated by \citet{morgan2007fcorona}. }
\label{fcorona}
\end{figure}

In the absence of additional information, we assume that the excess density in the tomography maps, averaged over latitude, is due to the F-corona. Figure \ref{fcorona} gives this mean value, along with error bars gained from the variance across latitude. This value is extrapolated out to large heights, beyond that measured (as shown by the solid line of figure \ref{fcorona}a), and distributed along a set of extended lines-of-sight in a spherically-symmetric distribution. Integrating, using appropriate geometric weights (e.g. \citet{quemerais2002}), gives an estimate of the polarized brightness of the F-corona, shown in figure \ref{fcorona}b. Comparing this polarized brightness with the total F-coronal brightness in polar regions given by \citet{morgan2007fcorona} gives the percentage polarization of the F-corona at heights between 4 and 5.5\Rs, shown in figure \ref{fcorona}c. This percentage is more or less constant with height, at $6.6\pm0.9$\%. This estimate is based on the assumption of a constant outflow speed. If the coronal hole fast solar wind is accelerating at these heights, this estimate is a lower limit, and the true polarized F-corona component will be higher. However, if a component of this mean excess density is due to a calibration issue, the true polarized F-corona component will be lower.

The tomography therefore provides a way of quantifying certain errors in the K-coronal polarized brightness observations. These will be investigated in more detail in future work, including the analysis of a larger dataset spanning different periods of the solar cycle. Hopefully this will lead to an improved calibration for the COR2A instrument.

\section{Distribution of streamers}
\label{results}

This section presents density maps from solar minimum and the ascending phase close to solar maximum, gained using the data calibration of \paperi, the tomographical method of \paperii, and the narrowing, refining and correcting methods described in section \ref{method}. Here we describe the coronal structures qualitatively, and show quantitatively how densities vary as a function of height within streamers and coronal holes in the following section.

\subsection{An equatorial streamer belt near solar minimum}
\begin{figure}[]
\begin{center}
\includegraphics[width=8.5cm]{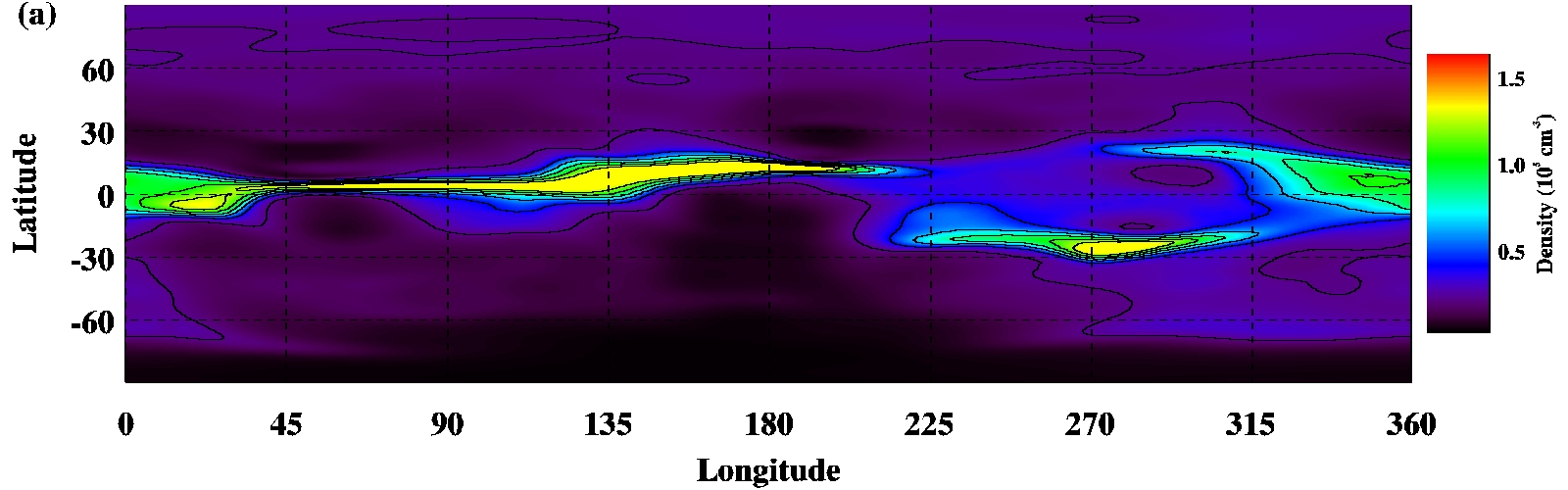}
\includegraphics[width=8.5cm]{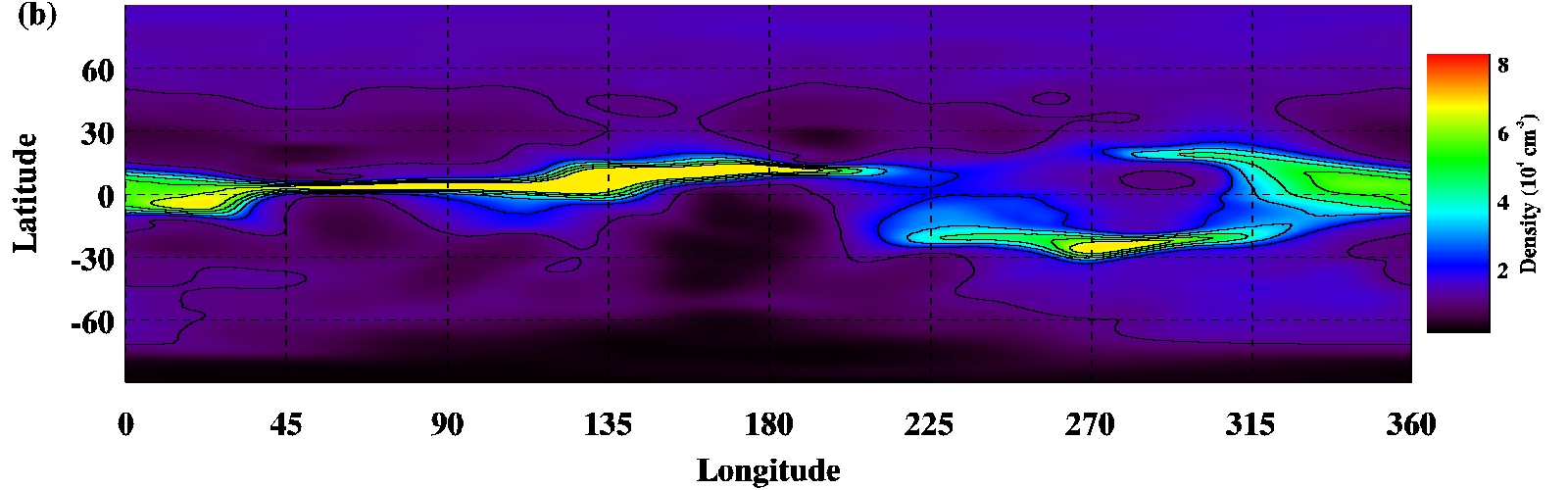}
\includegraphics[width=8.5cm]{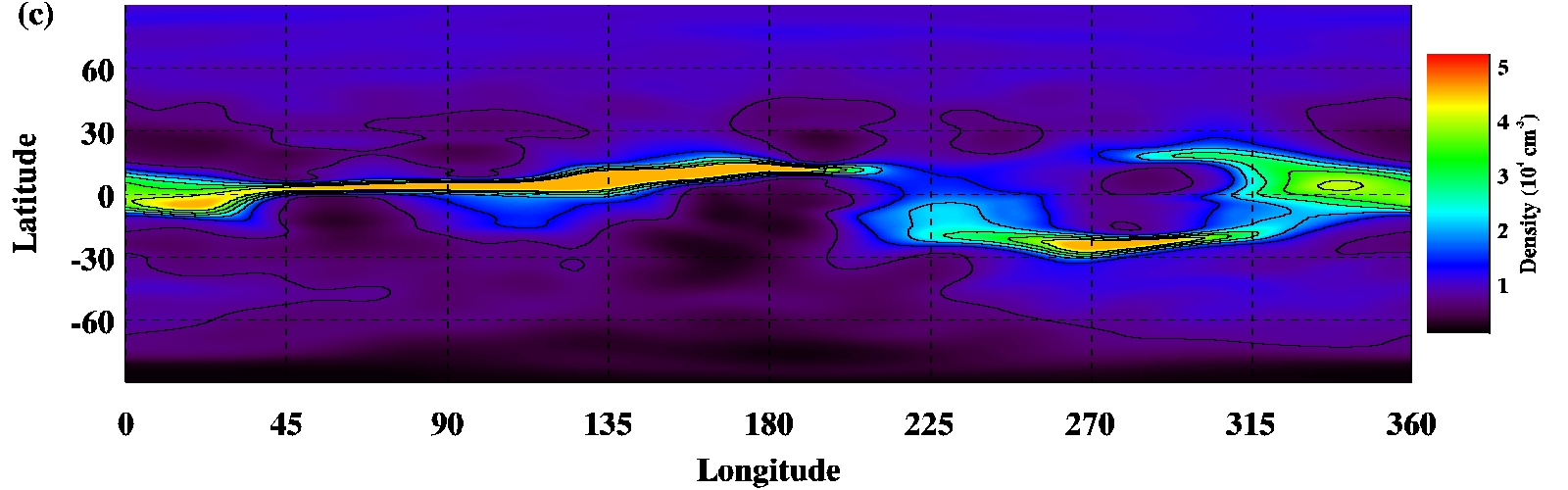}
\includegraphics[width=8.5cm]{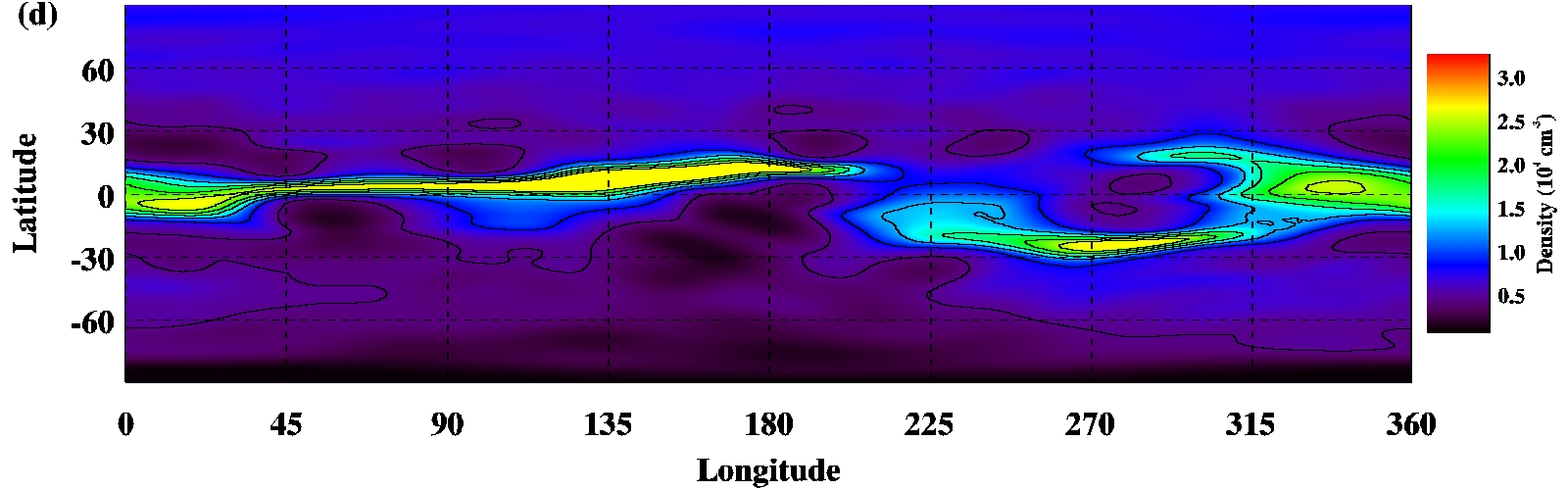}
\includegraphics[width=8.5cm]{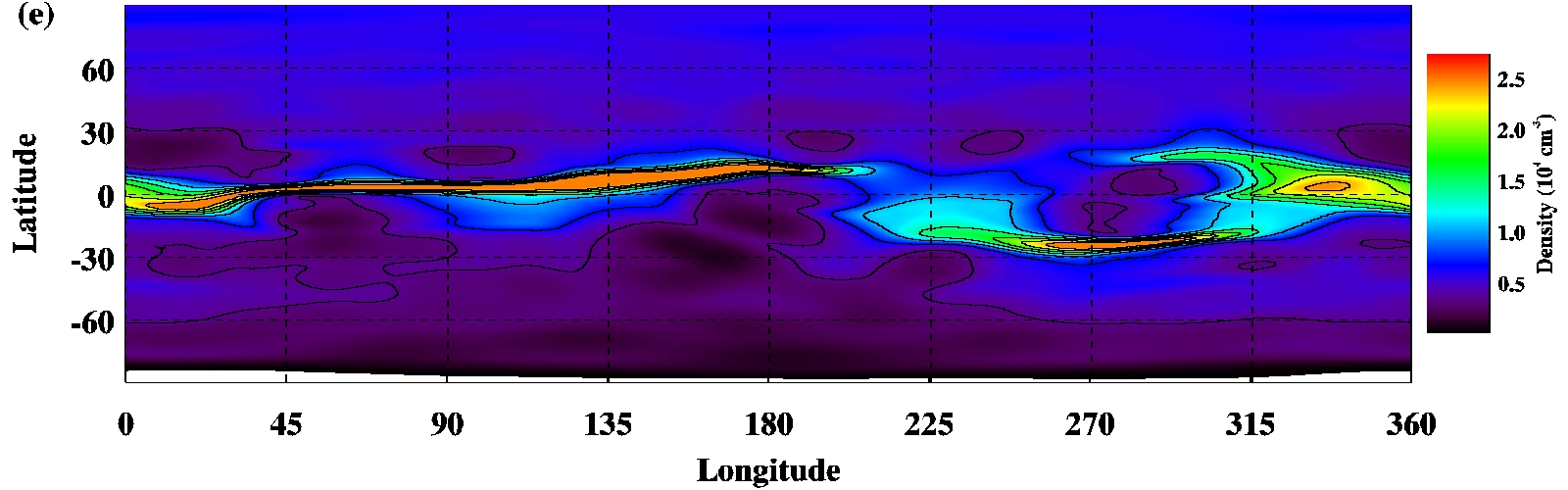}
\end{center}
\caption{Density maps of the solar corona for mid-date 2018/11/11 in Carrington longitude-latitude coordinates formed using the tomographical spherical harmonic approach of \paperii\ and this paper's method to narrow streamers at heights (a) 4.0\Rs, (b) 5.0\Rs, (c) 6.0\Rs, (c) 7.0\Rs, and (c) 8.0\Rs.}
\label{densmaps}
\end{figure}

Figure \ref{densmaps} shows reconstructed densities for middate 2018/11/11 at heights of 4 to 8\Rs. This is a period when the coronal structure is close to solar minimum, where the only deviation from a single narrow streamer sheet is a branching showing a streamer-pseudostreamer pair surrounding longitudes of 270\de. There are no significant differences in structure between heights, showing that the coronal structure is predominantely radial, as expected in the extended inner corona. The streamer belt can be as narrow as a few degrees of latitude in places, and is at most $\approx$15\de\ in width. The boundaries between streamers and surrounding regions are generally abrupt. Regions where there are smoother transitions from low to high density may in truth be sharper - the reconstruction process is expected to result in smoother structures than the true corona, particularly near the equator as discussed in \paperii. 

\begin{figure}[]
\begin{center}
\includegraphics[width=7.5cm]{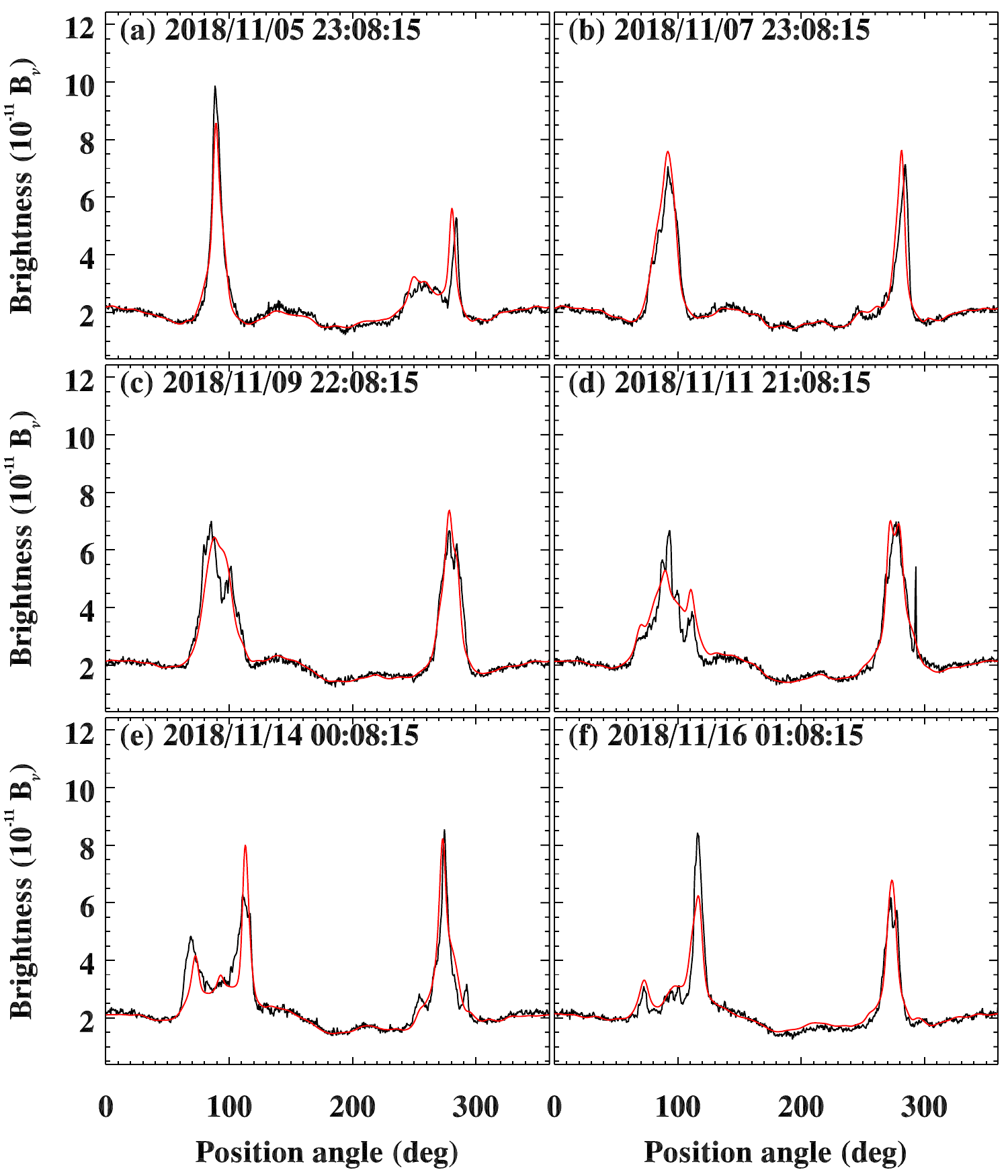}
\end{center}
\caption{Comparison of observed (black) and reconstructed (red) brightness, as a function of position angle, for several dates during the two-week observation period as indicated in each plot (a)-(f).}
\label{bobsmod2}
\end{figure}

In order to show the quality of this reconstruction, comparisons between observed and reconstructed brightnesses are shown in figure \ref{bobsmod2} at height 6.0\Rs\ for a few dates during the two-week observation period. The black lines give the observed brightness as a function of position angle. The red line shows the reconstructed brightness for the density map of figure \ref{densmaps}c. In low-brightness coronal holes, the agreement is excellent. At streamers, the fit is good at most times. Some faint structures are well-replicated (for example, the faint local peak of brightness near 70\de\ position angle for date 2018/11/16 shown in figure \ref{bobsmod2}f. At other places, the reconstructed brightness is not as sharply structured or as high in value as the observed brightnesses, for example the east streamer of 2018/11/10 shown in figure \ref{bobsmod2}c. 

\begin{figure}[]
\begin{center}
\includegraphics[width=7.5cm]{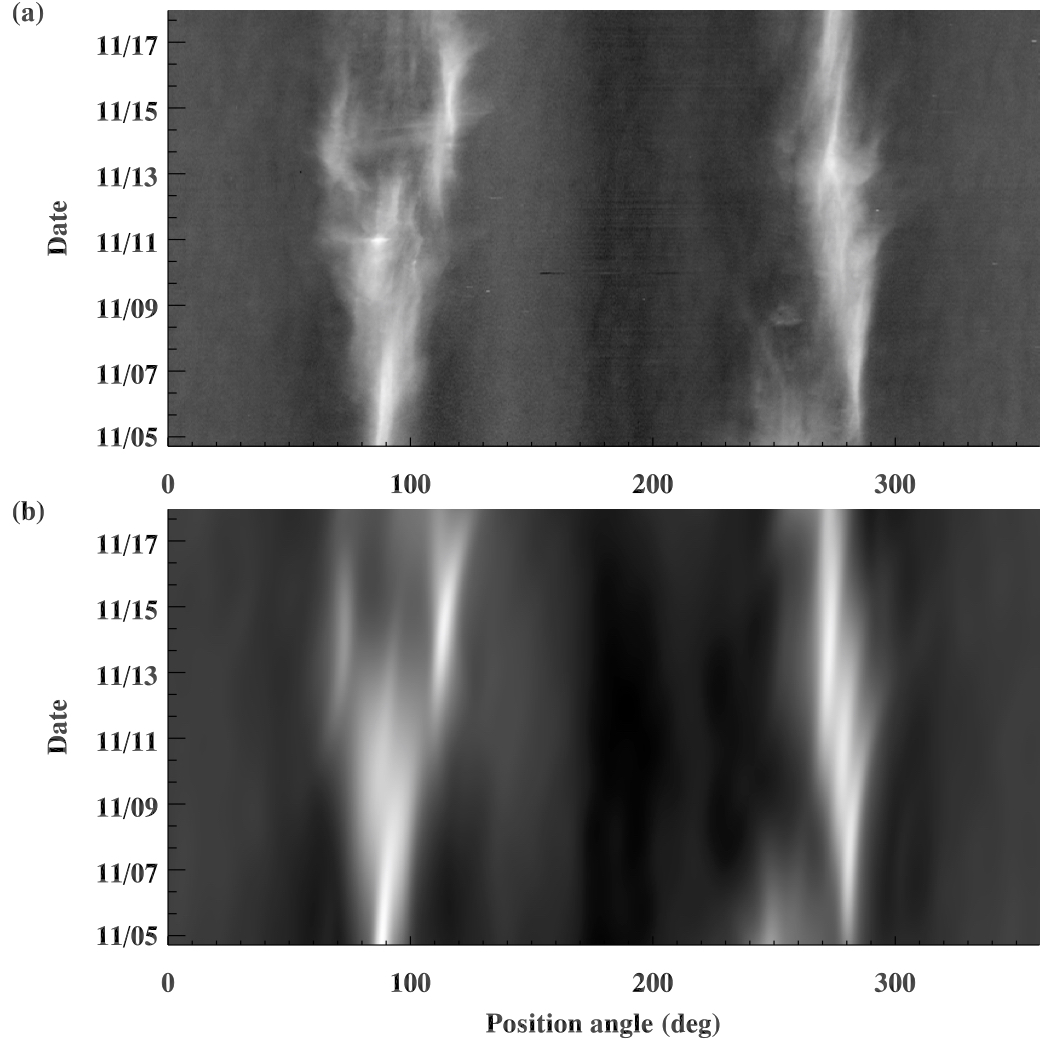}
\end{center}
\caption{Qualitative comparison of (a) observed and (b) reconstructed brightness, as a function of position angle and time, over the two-week observation period.}
\label{bobsmod}
\end{figure}

Figure \ref{bobsmod} shows a qualitative comparison between the whole set of observed and reconstructed brightness. It is apparent from this figure, and from figure \ref{bobsmod2}, that the streamers can contain very narrow structures that the tomography fails to reconstruct. Even the very narrow streamers seen in the reconstructed densities can fail to reach this level of fine-scale structure. This suggests strongly that the narrowest fine-scale structures seen within the streamer belt are (1) extremely narrow and (2) have lifetimes considerably shorter than the two-week observation period. Furthermore, it appears from comparing the observed and reconstructed brightness that the streamer belt is not uniform in density. The `fuzziness' of the observed streamers is not entirely due to noise, and strongly suggest a very fine-scale spatial variation in their density. This fine-scale structure has been explored using a non-tomographical approach by \citet{decraemer2019}, who found that a streamer sheet was composed of  fine ray-like structures with density contrast of a factor of 3 (see also \citet{thernisien2009}). Such a fine spatial structural variation to streamer density is also apparent from high-resolution eclipse images \citep[e.g.][]{habbal2014, alzate2017}. Very recently, results from the coronal imaging instrument Wide-field Imager for Solar PRobe (WISPR) aboard PSP is also revealing the fine ray-like structure of streamers \citep{poirier2020}.

Thus the smooth, largely uniform density of the streamer sheets shown in the tomography density maps is likely a local average of the true highly spatially variable, and probably time-variable, ray-like structure of the streamer belt. This means that the highest densities in streamers are considerably higher than those found using tomography, with the high densities limited to small spatial scales of a few degrees or less, and temporal scales of at most a few days.

\subsection{Streamers at higher latitudes near solar maximum}

\begin{figure}[]
\begin{center}
\includegraphics[width=8.5cm]{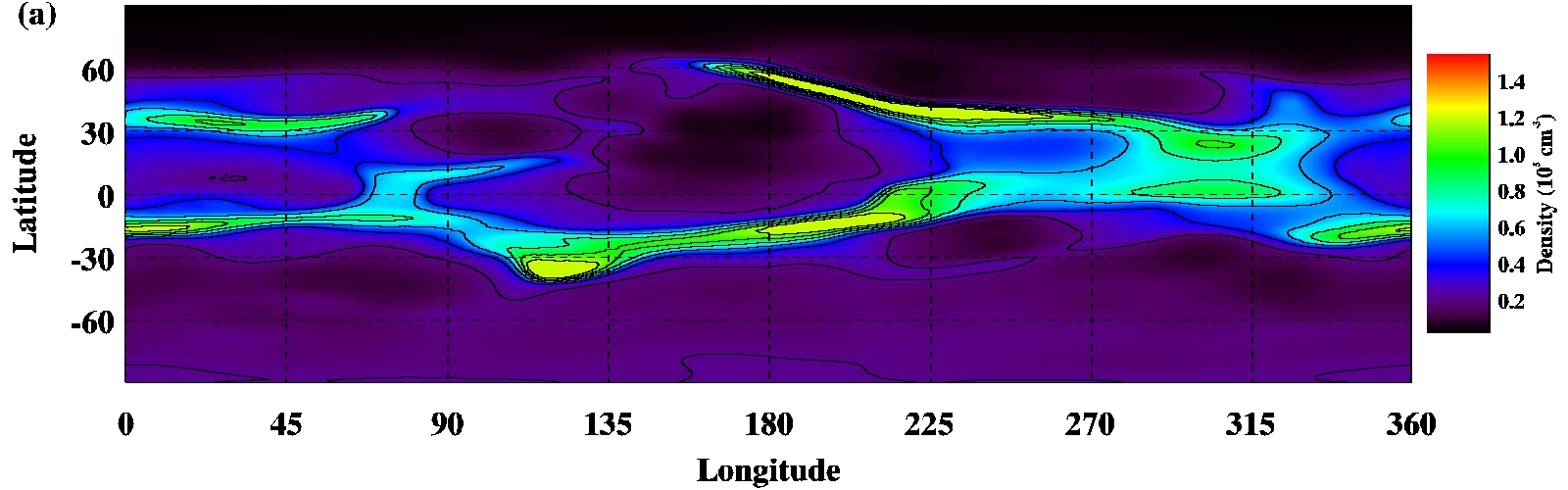}
\includegraphics[width=8.5cm]{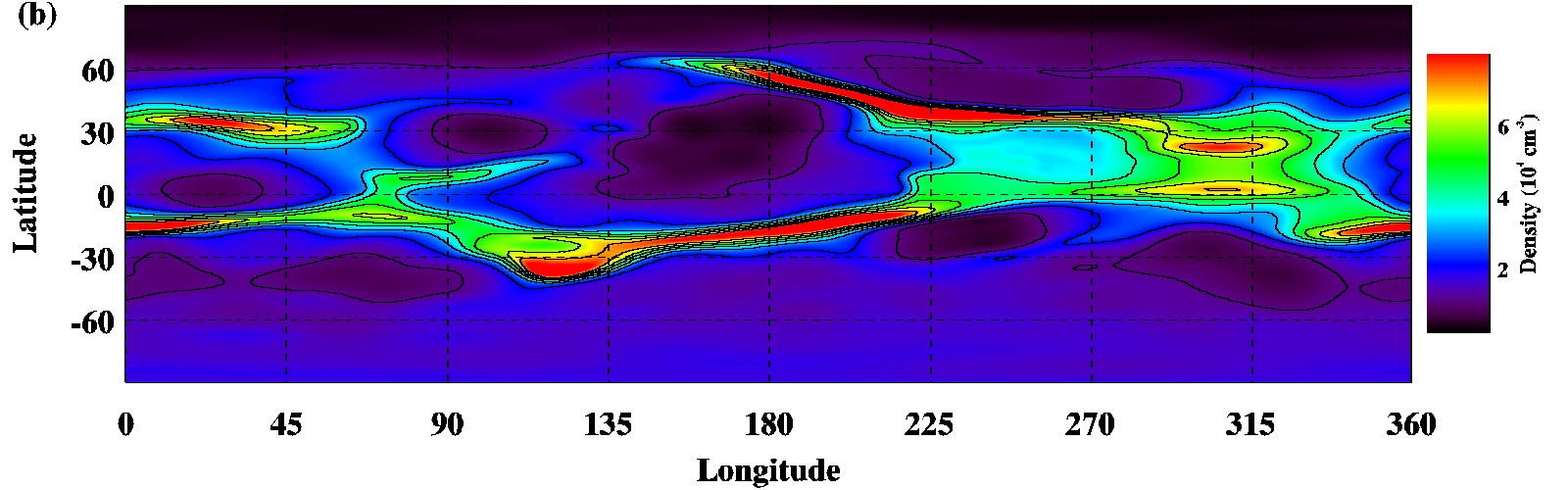}
\includegraphics[width=8.5cm]{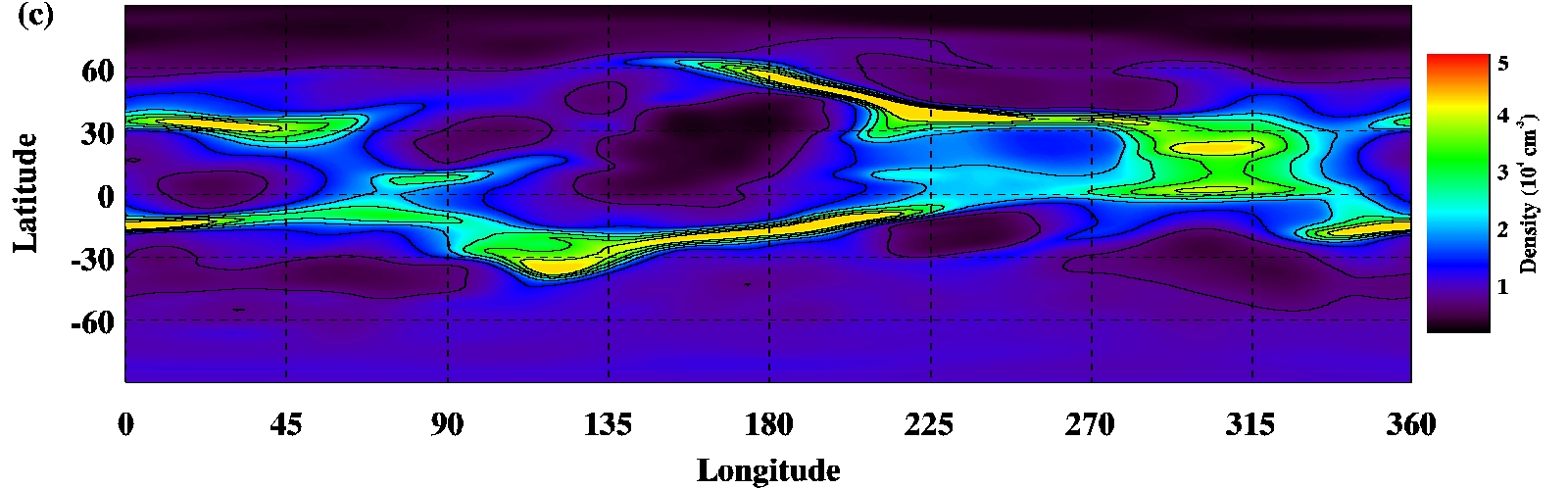}
\includegraphics[width=8.5cm]{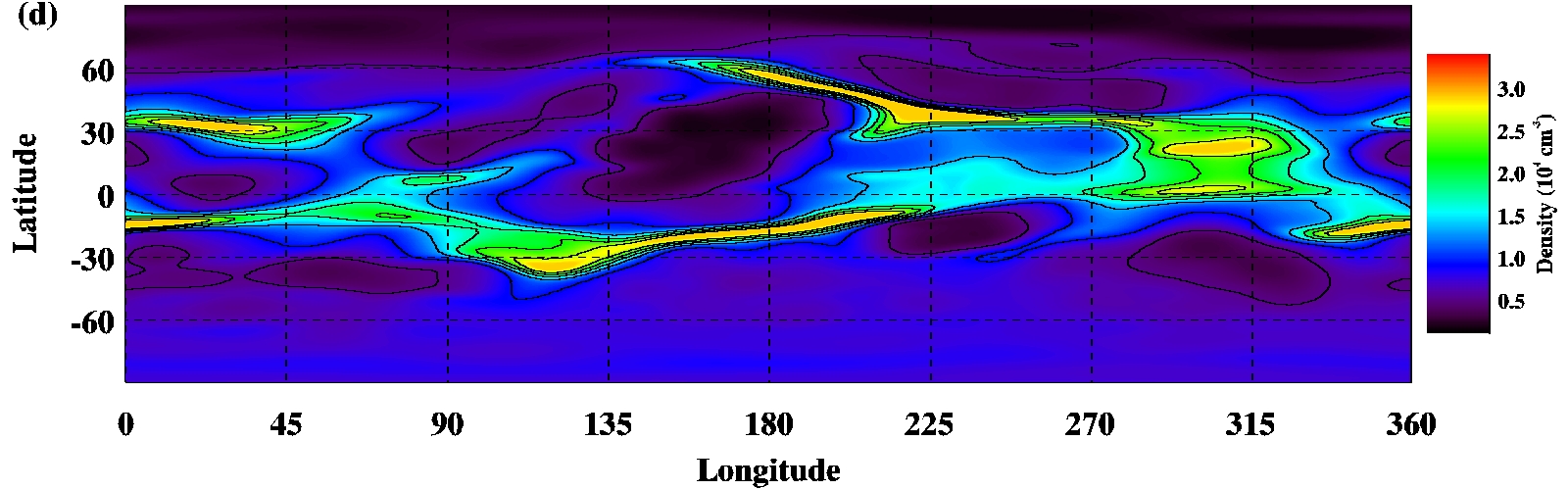}
\includegraphics[width=8.5cm]{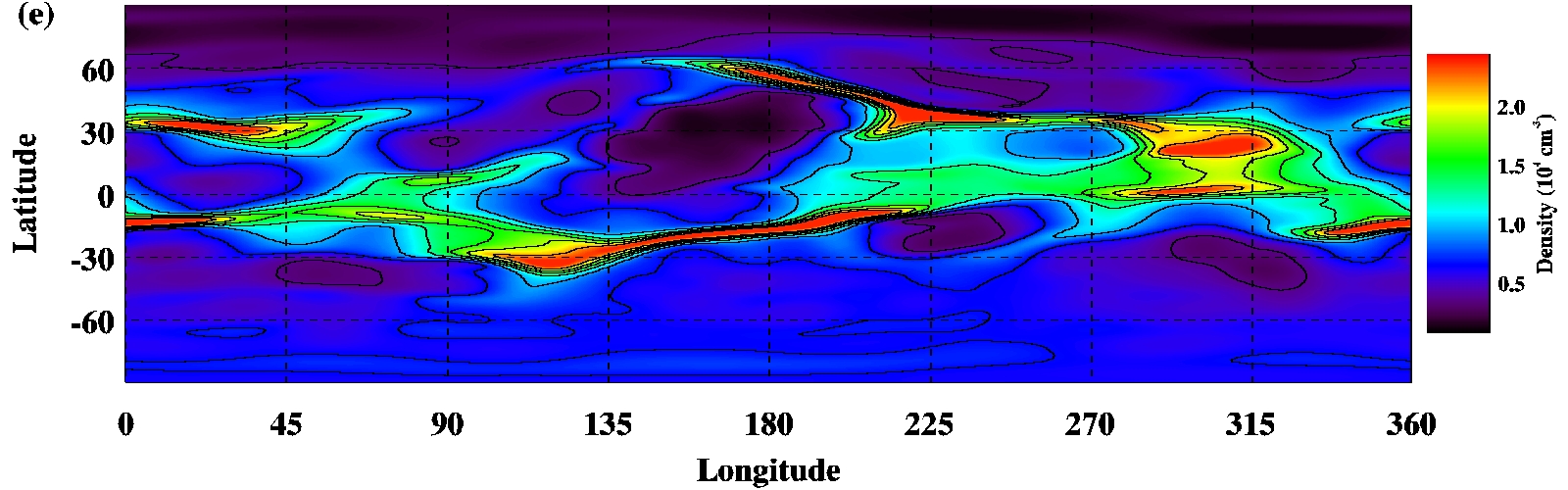}
\end{center}
\caption{Same as figure \ref{densmaps}, but for mid-date 2010/09/19 during the ascending phase to solar maximum.}
\label{densmapsmax}
\end{figure}

Figure \ref{densmapsmax}a-e show coronal densities for middate 2010/09/19, a period firmly in the ascending phase to solar maximum. Streamers are distributed at latitudes from -45\de\ to 60\de, with two narrow distinct bands at most longitudes that can be attributed to a global magnetic field dominated by a quadrapole. The approximate symmetric latitudinal variation of the two bands is not centered at the equator, but at around 10-20\de\ north. Between the streamer bands, there are large regions of low density, having densities similar to the densities of the polar coronal holes in some regions. The streamer densities are generally similar for both north and south bands. The streamer widths are narrow, similar to that found for solar minimum, ranging from a few to $\approx$15\de. Near longitudes 70\de\ and 315\de, the two streamer bands drift close in latitude, and may merge at 315\de.

\begin{figure}[]
\begin{center}
\includegraphics[width=7.5cm]{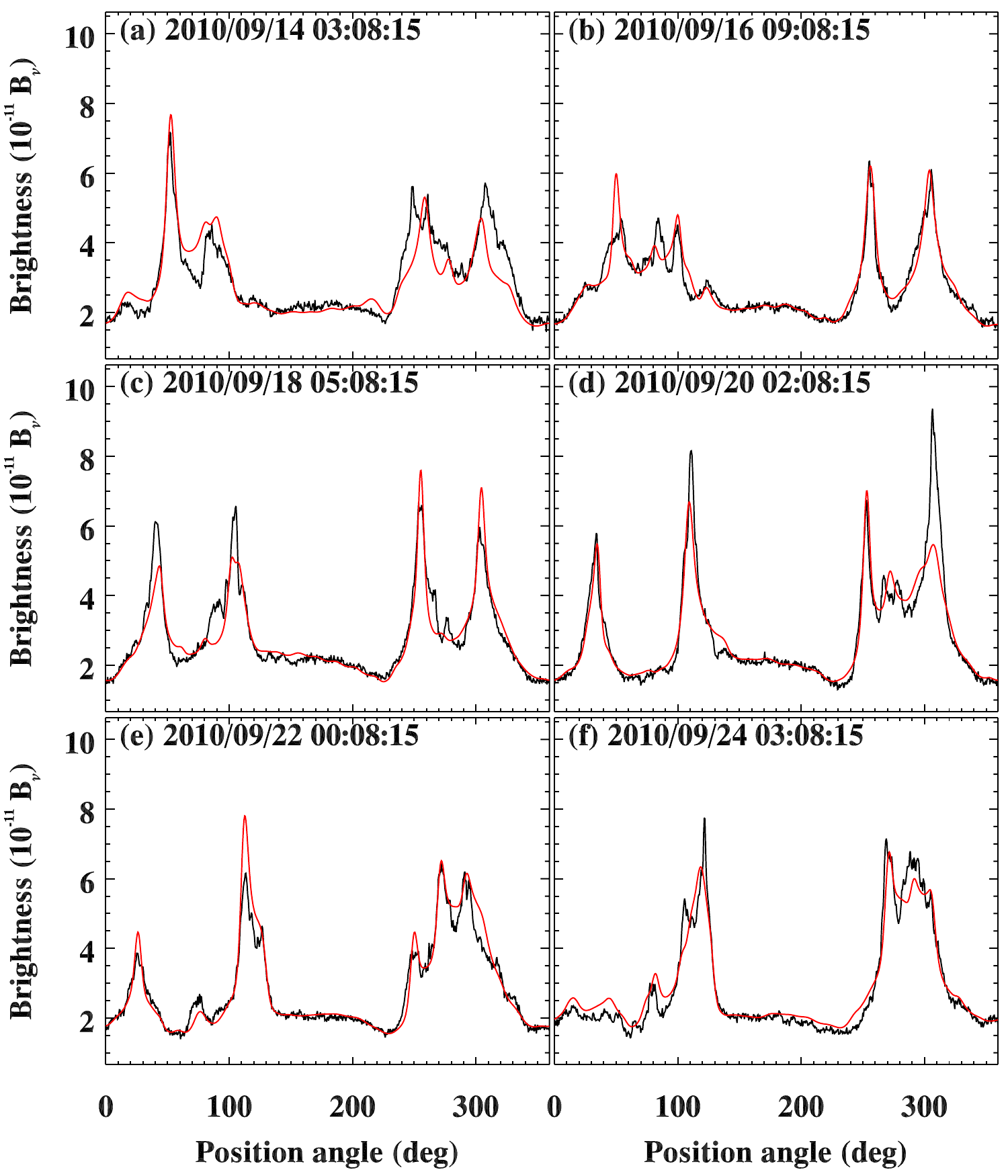}
\end{center}
\caption{Same as figure \ref{bobsmod2}, but for mid-date 2010/09/19 during the ascending phase to solar maximum.}
\label{bobsmod2max}
\end{figure}

In order to show the quality of this reconstruction, comparisons between observed and reconstructed brightnesses are shown in figure \ref{bobsmod2max} at height 6.0\Rs\ for a few dates during the two-week observation period. As with the solar minimum examples of figure \ref{bobsmod2}, in low-brightness coronal holes, the agreement is generally excellent. Considering the difficulty of reconstructing the solar maximum corona, the fit is good at most times and position angles, and is excellent for several narrow streamers. At other places, the reconstructed brightness is not as sharply structured or as high in value as the observed brightnesses. 

\begin{figure}[]
\begin{center}
\includegraphics[width=7.5cm]{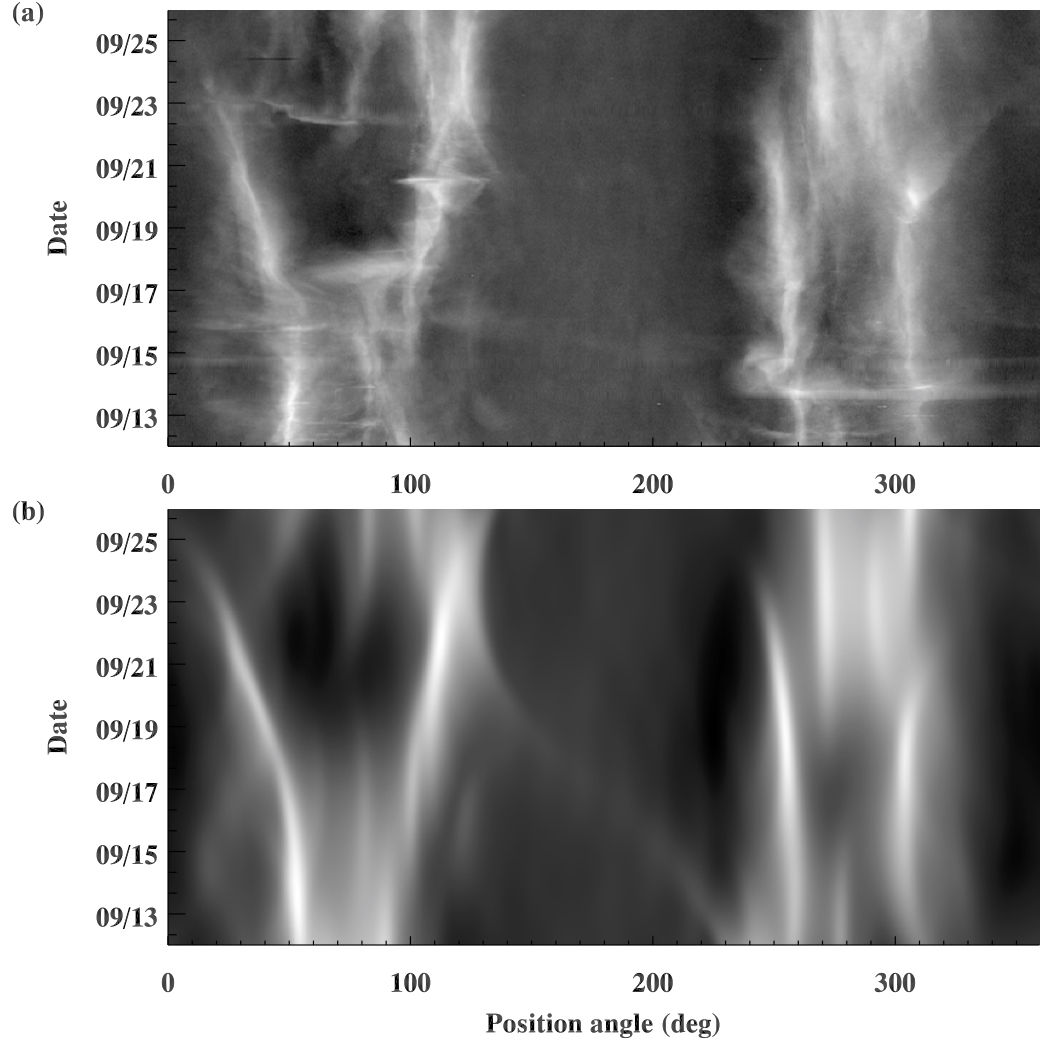}
\end{center}
\caption{Same as figure \ref{bobsmod}, but for mid-date 2010/09/19 during the ascending phase to solar maximum.}
\label{bobsmodmax}
\end{figure}

Figure \ref{bobsmodmax} shows a qualitative comparison of brightness. In contrast to the example from solar minimum, faint streaks of brightness are seen crossing the dim polar coronal holes, caused by streamer structures at high latitudes. The streamers are concentrated more to the north (the south polar coronal hole covers a far larger range of position angles than the north), confirming the northerly distribution of the dual streamer belt revealed by the tomography. The match between observed and reconstructed brightness is good, and shows that the method can reliably reconstruct the solar maximum corona. 

\section{Coronal densities}
\label{sectiondensity}
\begin{figure}[]
\begin{center}
\includegraphics[width=7.0cm]{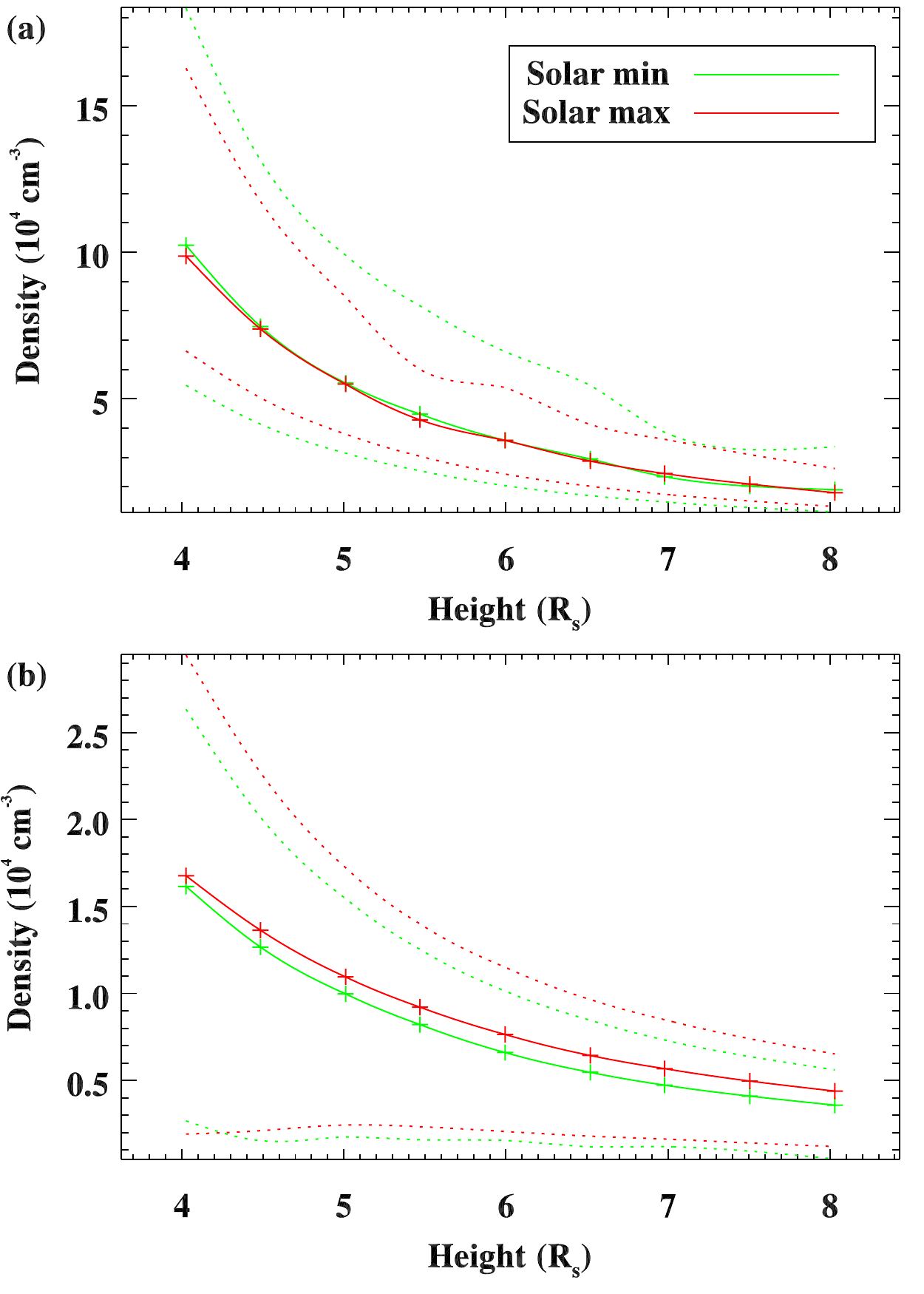}
\end{center}
\caption{Densities as a function of height for (a) streamers and (b) coronal holes. The solid green and red lines show the densities for solar minimum and maximum respectively. The mean values are calculated over regions where the density is greater than 1.9 times the mean density for streamers, and lower than the mean density for coronal holes. These thresholds are found by manual inspection of masked areas that effectively include or exclude streamers. The dotted lines show the $5^{th}$ percentile minimum and maximum densities.}
\label{density}
\end{figure}

Mean density values are calculated for a set of tomography maps for both the 2018/11/11 solar minimum and 2010/09/19 ascending phase at 0.5\Rs\ height increments between 4 and 8\Rs. This is done separately for streamer regions, defined as where the density is greater than 1.9 times the mean density, and for coronal holes, defined as where the density is lower than the mean density. These mean densities are shown for streamers in figure \ref{density}a and coronal holes in figure \ref{density}b as the unbroken green (red) lines for solar minimum (maximum). The dotted lines show the $5^{th}$ percentile minimum and maximum densities. An important finding is that mean streamer and coronal hole densities are very similar for the solar minimum and maximum periods. The streamer mean densities, in particular, are almost identical. Mean densities decrease from $10^5$\cmvol\ to $2 \times 10^4$\cmvol\ in streamers, and from $1.6 \times 10^4$\cmvol\ to $5 \times 10^3$\cmvol\ in coronal holes over this height range. Maximum densities in streamers decrease from approximately $1.6 \times 10^5$\cmvol\ to $3 \times 10^4$\cmvol. 

\begin{figure}[]
\begin{center}
\includegraphics[width=7.5cm]{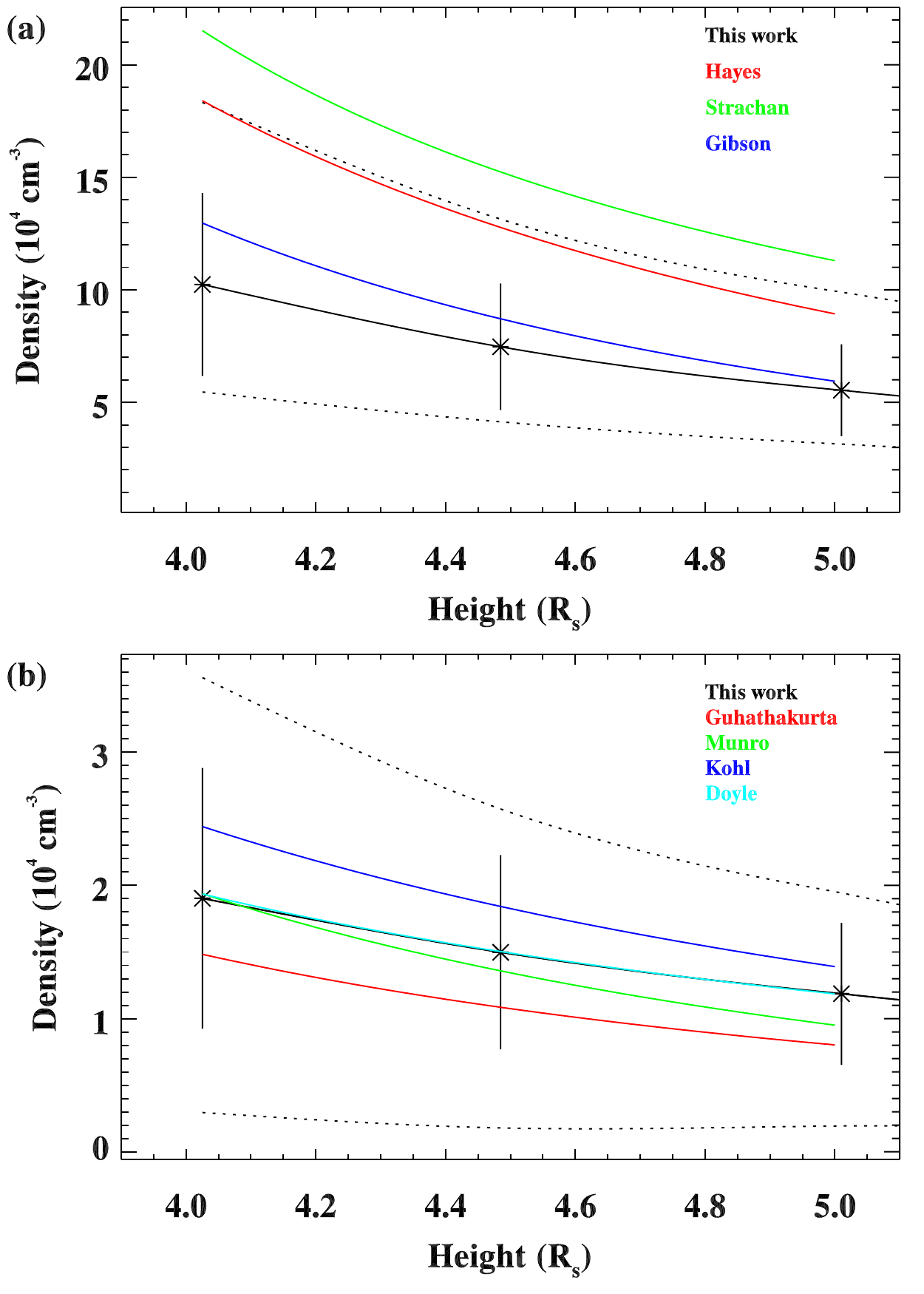}
\end{center}
\caption{Mean densities as a function of height for (a) streamers and (b) coronal holes. The black solid lines show the densities for solar minimum, and the dotted line show the $5^{th}$ percentile minimum and maximum. The error bars show the standard deviation of densities at each height. Published values are shown as coloured lines: (a) for streamers as red \citep{hayes2001}, green \citep{strachan2002}, blue \citep{gibson1999}; and (b) for coronal holes as red \citep{guhathakurta1999}, green \citep{munro1977}, blue \citep{kohl1998}, and cyan \citep{doyle1999}. Note that the studies of both \citet{gibson1999} and \citet{kohl1998} are limited to heights below 4\Rs, and their analytical functions of density have been extrapolated to 5.0\Rs\ here to enable comparison. \citet{hayes2001} and \citet{strachan2002} do not provide analytical functions of density, and the profiles shown here are gained from the parameter fitting to scanned plots of \citet{harding2019}.}
\label{densitylit}
\end{figure}

Figure \ref{densitylit} compares the solar minimum density to published values for (a) equatorial streamers and (b) polar coronal holes. For streamers, the mean tomographical densities are in good agreement with \citet{gibson1999}. \citet{hayes2001} and \citet{strachan2002} are at, or above, the upper range of tomographical densities. This is difficult to explain since a non-tomographical inversion depending on a prescribed geometrical distribution (e.g. axial or spherical symmetry) generally leads to an underestimate of streamer density. 

Mean coronal hole tomographical densities are in good agreement with the literature, with four published values falling within one standard deviation of the mean density at all heights. This gives confidence both to the published values based on inversions with prescribed symmetrical distribution, and to the density excess corrections applied to the tomography maps, described in section \ref{fcoronasec}.

\section{Streamer mass and outflow speeds}
\label{sectionmass}

\begin{figure}[]
\begin{center}
\includegraphics[width=7.5cm]{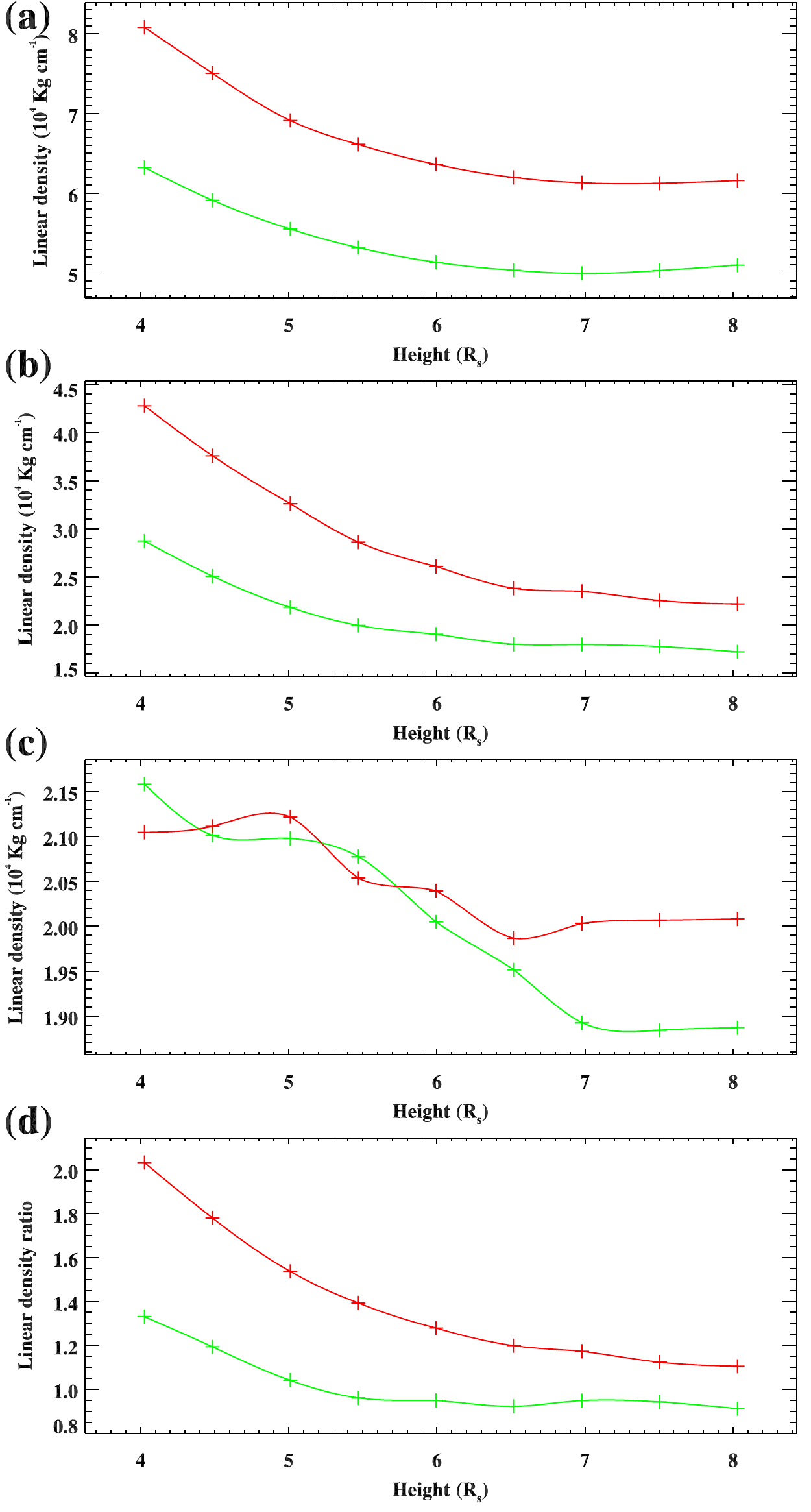}
\end{center}
\caption{(a) Electron density converted into mass density assuming 10\%\ Helium and integrated over a spherical surface area, giving a linear density at several heights, and equivalent to the plasma mass contained within a spherical shell of thickness 1cm. This is shown for solar minimum (green) and maximum (red), (b) linear density contained with streamers, (c) linear density within coronal holes, (d) the ratio of streamer to coronal hole linear density.}
\label{lineardensity}
\end{figure}

Assuming a 10\%\ particle fraction of Helium in the corona, the electron density can be converted into a mass density. This is integrated over the surface area of the sphere to give a linear density (equal to the total plasma mass contained within a spherical shell of 1cm thickness at a given height). This is shown as a function of height in figure \ref{lineardensity}a for solar minimum (green) and maximum (red). Despite the similarity of solar minimum/maximum densities, the solar maximum corona contains $\approx$30\%\ higher mass due to the greater volume occupied by streamers. In a radially-expanding corona with constant outflow velocity, the mass profiles should remain constant with height, whilst a decrease is seen here.

Figure \ref{lineardensity}b shows the mass contained within streamers. This mass decreases by almost half over the height range, showing that the slow solar wind within streamers is under acceleration. The solar maximum corona contains approximately 50\%\ greater mass within streamers at 4\Rs\ compared to solar minimum, with this ratio decreasing to 35\%\ at 6.5\Rs, suggesting an increased solar wind acceleration within streamers during the solar maximum period. The mass contained within coronal holes, shown in figure \ref{lineardensity}c is similar for both minimum and maximum periods. The fairly constant values with height is to be expected given the excess density correction described in section \ref{fcoronasec}. Figure \ref{lineardensity}d shows that the ratio of streamer to coronal hole mass is decreasing with height. At heights of 5.5\Rs\ and above, the solar minimum corona contains roughly equal mass in streamers and coronal holes, whilst the solar maximum corona contains twice as much mass in streamers than coronal holes at 4.0\Rs\, a ratio which drops close to one by 8\Rs. This is due to the larger area occupied by streamers during solar maximum. The decreasing ratio is due to the outflow acceleration at extended heights in streamers compared to coronal holes.

\begin{figure}[]
\begin{center}
\includegraphics[width=7.5cm]{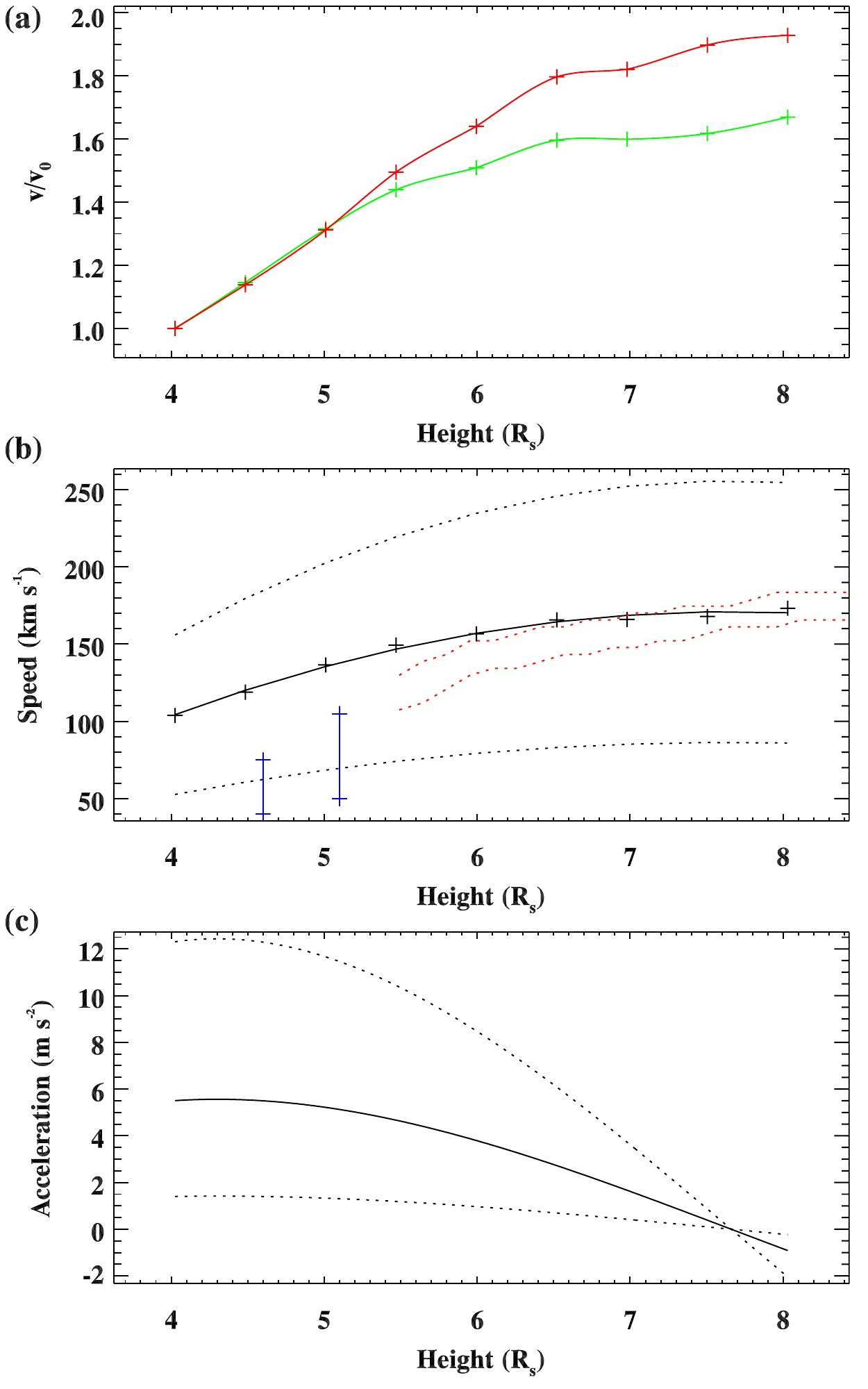}
\end{center}
\caption{(a) The ratio of streamer linear densities to the base streamer linear density at 4\Rs\ shown for solar minimum (green) and maximum (red). This ratio is equal to the outflow velocity ratio given the radial expansion of streamers at these heights. (b) Central (solid black line), and lower and upper estimates (dotted black lines) of outflow velocity given the constraint of mean proton flux measured by PSP. The black crosses give values calculated directly from the velocity ratio of (a), and the lines give a second order polynomial fit to height. The blue points and ranges are as given by \citet{frazin2003}, estimated from Doppler dimming diagnostics of O$^{5+}$ emission lines measured by UVCS/SOHO. The dotted red lines are constraints estimated by \citet{cho2018} using Fourier filtering of LASCO C3/SOHO data from 2009. (c) The solar wind acceleration based on the fitted velocities of part (b).}
\label{acceleration}
\end{figure}

Figure \ref{acceleration}a shows the ratio of the streamer linear density (shown in figure \ref{lineardensity}b) over height to the base linear density at 4\Rs. Given the radial expansion of the streamer structure as seen in the tomography maps, this ratio is equal to the ratio of the outflow velocity over the base outflow velocity at 4\Rs. For solar maximum (red line), the acceleration is constant between 4 and 6.5\Rs, then starts to decrease. The outflow velocity at 8\Rs\ is close to double that at 4\Rs. For the solar minimum corona, the acceleration starts to decrease at around 5\Rs, and the outflow velocity at 8\Rs\ is approximately 1.6 that at 4\Rs.

\begin{figure*}[]
\begin{center}
\includegraphics[width=16.5cm]{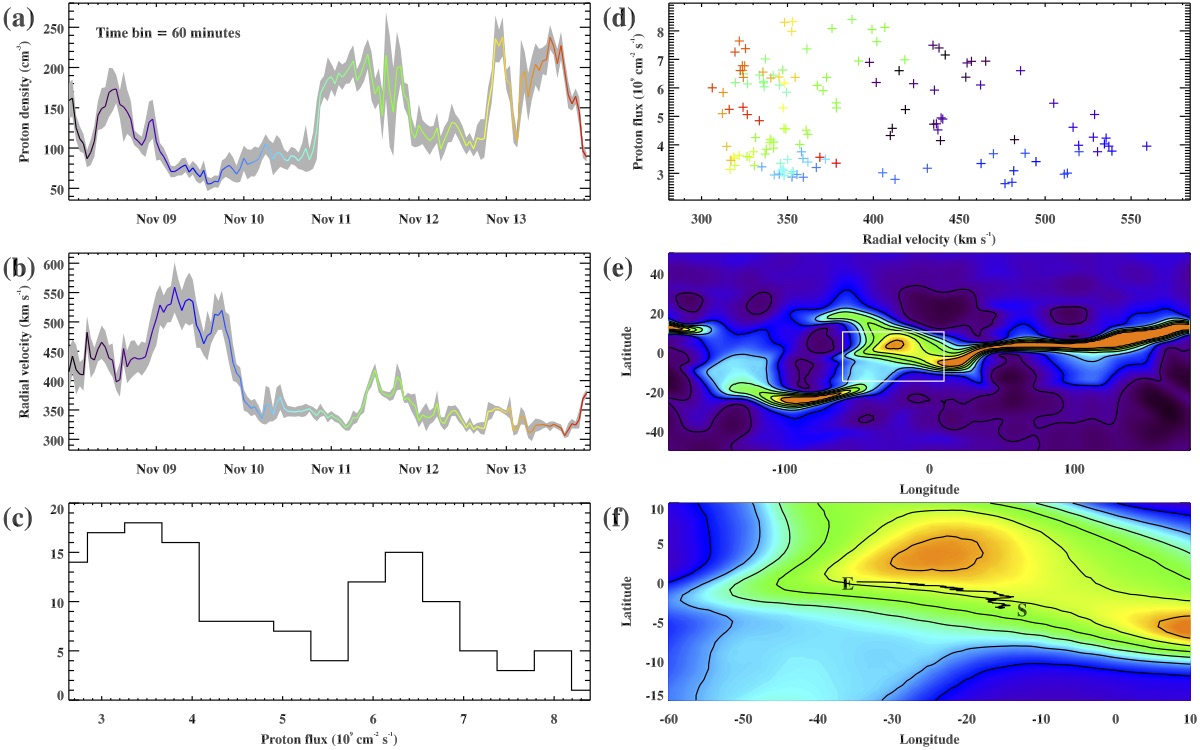}
\end{center}
\caption{(a) Proton density measured by PSP during the November 2018 perihelion. The measurements are averaged over 1-hour periods. The grey shading shows the standard deviation of values over the 1-hour bins. The mean density values are shown with a coloured line, with colour changing with a rainbow colour palete from purple to red over time. (b) Proton mean radial velocity measured by PSP. (c) Histogram of proton flux measured by PSP over this period. Each histogram number corresponds to a 1-hour bin. (d) Proton flux as a function of radial velocity. The colours correspond to time, as shown in (a) and (b). (e) Tomography map for a height of 8\Rs\ and mid-date 2018/11/11. The boxed white region is shown in more detail in (f). The black path shows the path of PSP, ballistically extrapolated down to the tomography map height using the measured radial velocities. Label S is the start of the path (2018/11/08 00:00) and E is the end (2018/11/14 00:00). }
\label{psp}
\end{figure*}

An independent measure of mass flux within streamers allows an estimate of outflow speed, based on the tomography map densities and the assumption of constant mass flux. During November 2018 the Parker Solar Probe (PSP) was at perihelion. The Solar Wind Electrons Alphas and Protons (SWEAP) instrument sampled proton velocity distributions. The proton density, averaged over 1 hour periods between November 8 to 13, is shown in figure \ref{psp}a, and the mean radial velocity in figure \ref{psp}b. At the start of this period, PSP samples a high-speed stream ($v_r \approx 500$\kms), then a slow stream ($v_r \approx 350$\kms). Densities vary greatly, despite the 1-hour averaging. Figure \ref{psp}c shows a histogram of proton particle flux over this 6-day period, and figure \ref{psp}d shows the flux as a function of radial velocity. This figure shows that there is no particular dependence of proton flux with radial velocity, although there are fewer high flux values in the high-speed stream. 

Figure \ref{psp}e shows a tomography map for a height of 8\Rs, with the white boxed region shown in greater detail in figure \ref{psp}f. The path of PSP, ballistically extrapolated down to the tomography map height using the measured radial velocities, is shown in figure \ref{psp}f and is labelled with letter S for start (2018/11/08 00:00) and E for end (2018/11/14 00:00). The path is close to the center of the streamer belt, and the tomography indicates fairly uniform density, in contrast to the PSP measurements. This leads to two conclusions: (1) The high-speed stream measured by PSP at the start of the period is originating very close to the streamer belt, suggesting that the high-density streamer shown in the tomography map should be considerably narrower, so that the PSP path moves from a high to low-speed region. Either this, or there may be high-speed regions within the higher-density boundaries of the streamer belt. (2) The density shown in the tomography maps is far smoother than the true corona, as discussed earlier. The high variation of densities measured by PSP demand a far higher variation than that estimated by the tomography.

The PSP mean proton flux over this period can be used to estimate the mean outflow velocity within streamers. This is shown as the crosses in figure \ref{acceleration}b, with the solid line giving a second-order polynomial fit to the velocities as a function of height. The standard deviation of PSP-measured mass flux give a large range of outflow velocities, indicated by the dotted lines. We estimate mean outflow of 100\kms\ (range 50-150\kms) at 4\Rs, increasing to 170\kms\ (range 90-250\kms) at 8\Rs. Note that PSP is sampling over a small section of the streamer belt, whilst the tomography mass flux uses densities averaged over all streamer regions.  

In a study based on Doppler dimming diagnostics of O$^{5+}$ emission lines measured by the UltraViolet Coronagraph Spectrometer (UVCS, \citet{kohl1995}) aboard the Solar and Heliospheric Observatory (SOHO), \citet{habbal1997} showed that outflow speeds within solar minimum streamers at heights below 5.5\Rs\ remained below $\approx$94\kms. In a more detailed study of the same solar minimum streamers, \citet{frazin2003} gave possible ranges for outflow velocity at several heights below 5.5\Rs. Two of the constraints set by \citet{frazin2003} at 4.6 and 5.1\Rs\ are shown in figure \ref{acceleration}b as the blue points. They are at the lower limit of the PSP-constrained range. \citet{cho2018} applied a Fourier filtering technique to LASCO/SOHO C3 coronagraph data to estimate outflow speeds. Their estimate for equatorial regions, averaged over year 2009, is shown as the red dotted lines. Their estimates show larger acceleration, otherwise good general agreement, between heights of 5.5 and 8\Rs. 

From the fitted outflow velocities, acceleration is derived and shown in figure \ref{acceleration}c. The slow solar wind is proving a mean acceleration of approximately 6\mss\ (range 2-12\mss) at heights of 4 to 6\Rs, decreasing to zero below 8\Rs. The slow wind must accelerate to speeds of approximately 350\kms\ by distances of around 40\Rs\ in order to satisfy PSP measurements: even a very small residual acceleration of around 2\mss\ can achieve this from a starting speed of 160\kms\ at 8\Rs, thus our estimate of outflow speed is decreasing too rapidly at larger heights. As shown in figure \ref{acceleration}a, acceleration at solar minimum is lower than maximum, and decreases more rapidly at lower heights compared to maximum. This significant difference in acceleration between solar minimum and maximum is surprising given the similarity of mean streamer densities shown in figure \ref{density}a. Using the solar maximum velocity ratio of figure \ref{acceleration}a results in higher outflow speeds at 8\Rs, and a positive acceleration showing closer agreement to the profile of \citet{cho2018}.


\section{Visualising the tomography maps and data}
\label{sectionmap}

\begin{figure*}[]
\begin{center}
\includegraphics[width=10.0cm]{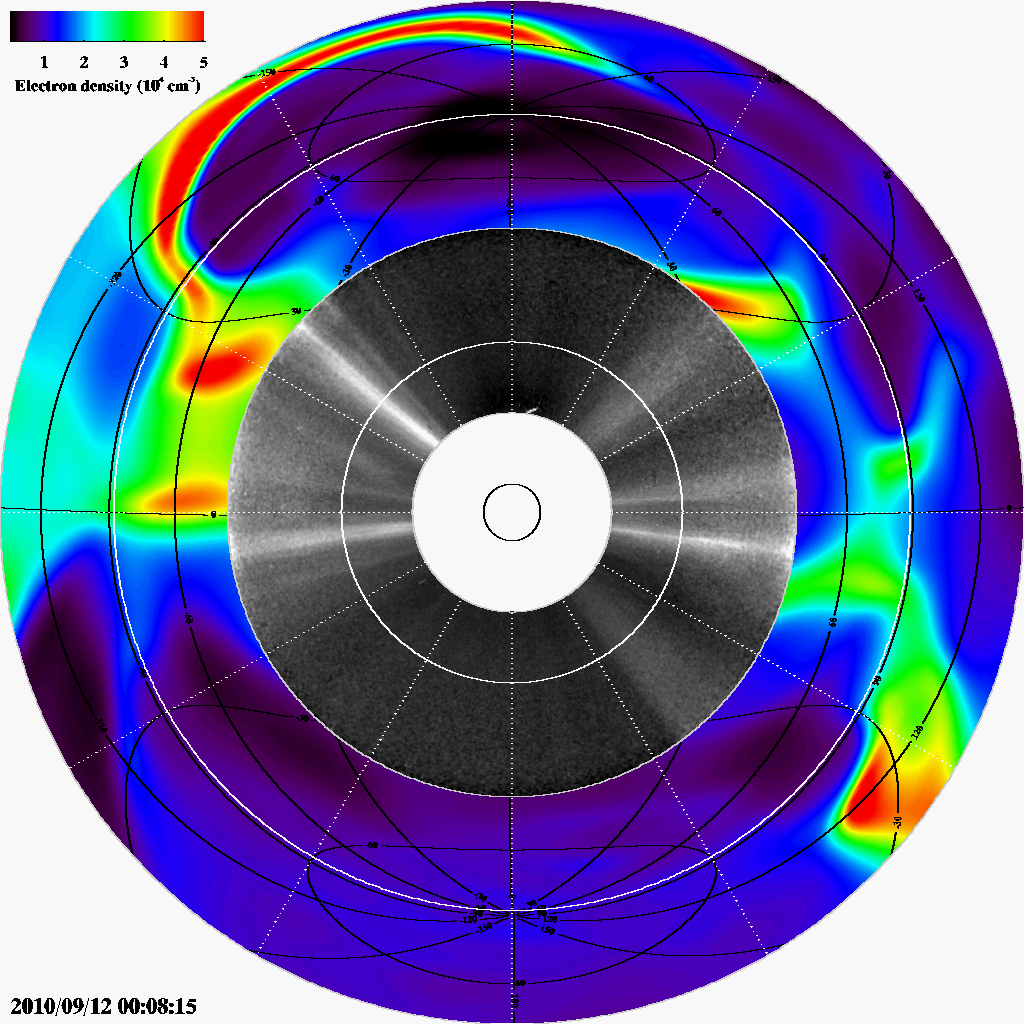}
\end{center}
\caption{The inner circular panel shows the COR2A polarized brightness observations processed using the Normalizing Radial Graded Filter \citep{morgan2006}. The height of interest is 6\Rs, shown with a solid white line approximately in the center of the field of view. The outer annular color panel shows the density along the line of sight for the height of interest, with the central solid white line denoting the points along the line of sight that are closest to the Sun (i.e. at a height of 6\Rs, and having the greatest influence on the observed brightness). The line of sight extends over ranges from $+$6\Rs\ (outer circle, farthest from observer) to $-$6\Rs\ (inner circle, closest to observer) from this point of closest approach. The black inner circle shows the position of the Sun. Carrington longitudes and latitudes are contoured as black solid lines over the colour density portion of the plot. A movie spanning two weeks of observations from 2010/09/12 to 2010/09/25 is available online.}
\label{tomo_movie}
\end{figure*}

As in the results shown above, the density maps are usually displayed on a Carrington longitude-latitude coordinate scale that is convenient for analysis and for comparison with other synoptic-type solar data (e.g. potential-field source surface magnetic field maps). They are not directly useful for comparison to the original coronagraph observations - this involves knowing the position of the east and west limbs, which are curved lines on the tomography maps, and move across the map with time. Figure \ref{tomo_movie} presents a way to more directly compare the density distribution with the observed brightness. 

In this mapping, the densities along lines of sight at each position angle and a single height of interest (6\Rs\ in this example), are interpolated from the Carrington density map. These are then displayed on a outer circular band surrounding the coronagraph observation. The central radius of this circular band corresponds to the point along the line of sight closest to the Sun. Points at a greater (lesser) radius are further (closer) along the line of sight from the observer. Thus, as this mapping is advanced in time (see online movie), one can see density structures moving inwards (outwards) at the east (west) corona, and crossing the point of closest approach to the Sun. This can be directly compared visually with the distribution of streamers in the coronagraph images. This is a very useful mapping for interpreting the observed brightness distribution, particularly for the complicated solar maximum corona, and can be easily extended for use on other types of data, most notably magnetic field extrapolations.

\section{Summary and future work}
\label{conclusions}
A new coronal tomography method, based on spherical harmonics and regularized least-squares inversion gives a robust reconstruction of the coronal density structure. Initially, the reconstructed streamers are overly wide due to the smoothing requirement of the regularization. This is remedied by a process that narrows all structure, with an optimal degree of narrowing found by a fit to the data. 

Applied to both solar minimum and maximum data, we show that:
\begin{enumerate}
\item To best match the data, coronal streamers must be very narrow, less than a few degrees in width in some regions.
\item The coronal structure is confirmed to be highly radial between heights of 4 and 8\Rs.
\item The mean densities of streamers and coronal holes are very similar between solar minimum and maximum.
\item For the two periods studied here, the solar maximum corona contains approximately 33\%\ more mass due to the larger area occupied by streamers. Streamers contain around 50\%\ more mass at solar maximum compared to solar minimum due to their larger area.
\item The decrease of streamer mass with height, particularly between 4 and 6\Rs, shows that the streamer slow solar wind is under considerable acceleration. This mass profile decreases only slowly above 6\Rs.
\item Based on PSP proton flux measurements, we approximately estimate that the slow solar wind speed accelerates from 100 to 170\kms\ between 4 and 8\Rs\ at solar minimum, in good general agreement with a previous study based directly on coronagraph images \citep{cho2018}.  
\item The solar maximum slow wind has a higher acceleration than solar minimum, and values between 1 to 10\mss\ are found for the slow wind at heights of 4 to 6\Rs. Above this height, acceleration decreases, but must remain at a mean of  approximately 1 to 2\mss\ between 8 and 40\Rs\ to satisfy PSP measurements.
\end{enumerate}

The very narrow reconstructed streamers give synthetic brightnesses that agree very well with the observed brightness (low residual). Despite this, some of the streamer substructures and fine, intricate variation seen in the data strongly suggest that streamers are composed of rays which are very narrow, and may have lifetimes considerably shorter than the half-solar-rotation used to form the reconstructions. This model is supported by previous studies \citep{morgan2007twistingsheets,morgan2007empiricalmodel,thernisien2009,decraemer2019}, by high-resolution eclipse observations \citep{habbal2014,alzate2017}, and by PSP measurements in this work and others \citep{howard2019}. A highly variable streamer belt must exist to explain the small-scale magnetic and density variations in slow wind measured by PSP \citep[e.g.][]{bale2019}, and other spacecraft further from the Sun. Coronal tomography cannot directly show such fine spatial and temporal scales.

The COR2A dataset, from 2007 to present, is currently being processed to provide a large set of tomography maps at two day increments over the height range of 4 to 8\Rs. These will be published, along with general results, as an atlas of the solar corona. For the first time, these maps can offer a crucial observational constraint on the distribution of slow and fast-wind structure in the extended solar corona and will help validate or improve global magnetic field extrapolations. They are valuable for interpreting remote observations of the corona, and for providing context for \emph{in situ} measurements, particularly for the close encounters of the Parker Solar Probe and Solar Orbiter. The tomography approach will also be a key process for maximizing the scientific value of coronagraph data from future missions, including the Polarimeter to Unify the Corona and Heliosphere (PUNCH) mission.

 \begin{acknowledgements}
We acknowledge (1) STFC grant ST/S000518/1 to Aberystwyth University which made this work possible, (2) the excellent facilities and support of SuperComputing Wales, and (3) the NASA Parker Solar Probe Mission and SWEAP team led by J. Kasper for use of data. The STEREO/SECCHI project is an international consortium of the Naval Research Laboratory (USA), Lockheed Martin Solar and Astrophysics Lab (USA), NASA Goddard Space Flight Center (USA), Rutherford Appleton Laboratory (UK), University of Birmingham (UK), Max-Planck-Institut fu\"r Sonnen-systemforschung (Germany), Centre Spatial de Liege (Belgium), Institut dÕOptique Th\'eorique et Appliq\'uee (France), and Institut d'Astrophysique Spatiale (France). 
 \end{acknowledgements}

\appendix
\section{Refining the density estimate}
\label{app1}
\begin{figure}[]
\begin{center}
\includegraphics[width=6.0cm]{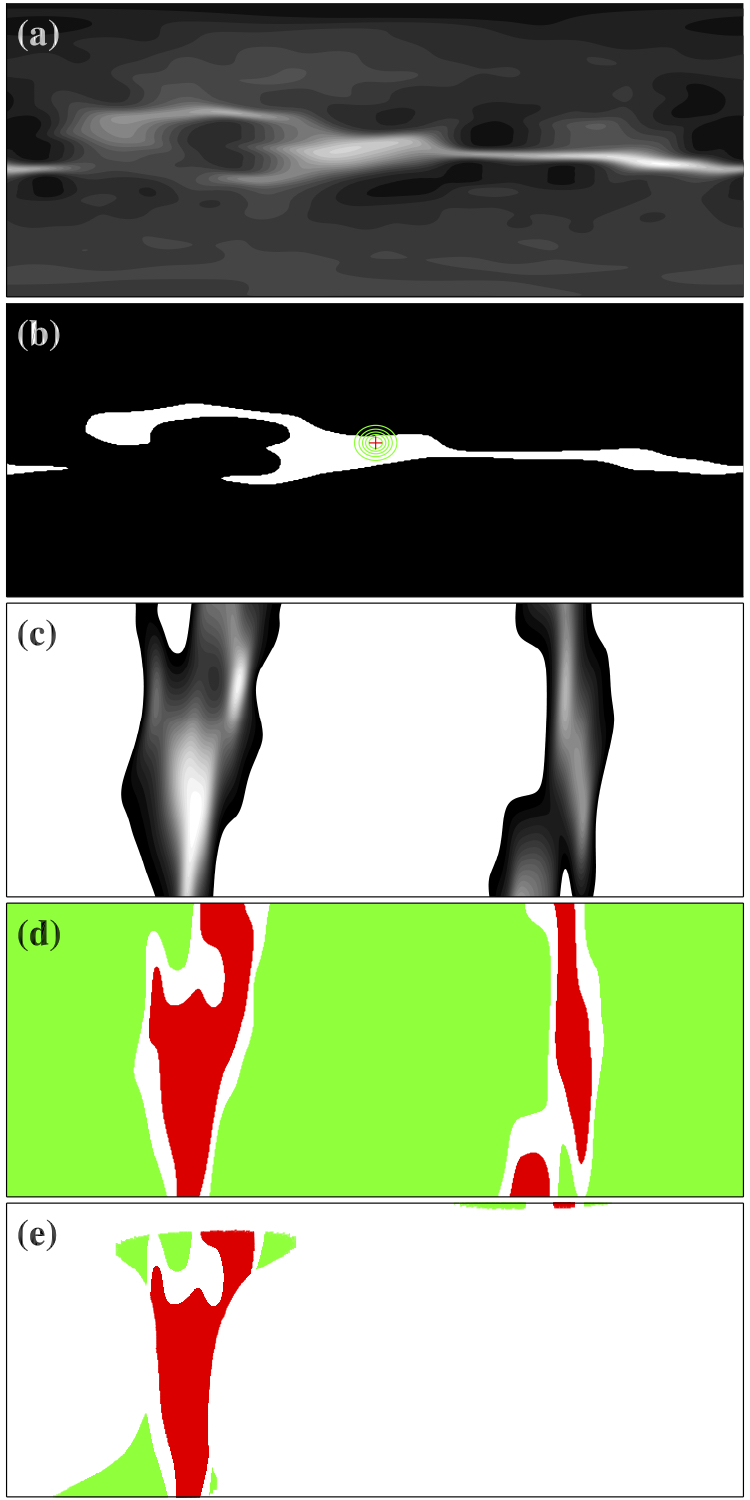}
\end{center}
\caption{(a) Narrowed density structure in longitude-latitude space. This density corresponds to the map shown in figure \ref{densnarrow}c for $f=200$. (b) Mask identifying position of streamers as described in the text. The green contours show the position of a Gaussian function centered on the red cross. (c) A mapping of the influence of streamer regions gained from line-of-sight integrations of the density mask shown in (b), in position-angle, time space. (d) Regions of the brightness identified as streamers (red) and coronal holes (green). (e) The red (green) points show streamer (coronal hole) regions that are influenced by the Gaussian function shown in green in (b).}
\label{refine}
\end{figure}

Figure \ref{refine}a shows the density narrowed with a factor of $f=200$. This case will be used as an example of the density refining method. A streamer mask is defined as regions of the density map that are greater than 1.7 times the mean density, shown in figure \ref{refine}b. This mask has values of 1 where streamers lie, and 0 elsewhere. A set of points are defined on a regular $8 \times 8$\de\ grid.  A Gaussian function is centered on each point, of width (standard deviation) 8\de. One of these points and functions is shown in figure \ref{refine}b as the red cross and green contours respectively. The Gaussian function values, $W_{\rho}$ will be used as a weighting later. Line-of-sight integrations of the streamer mask map into the position-angle, time space of the observation, shown in figure \ref{refine}c. High values in this array show where observations are strongly influenced by streamers. Broad regions have zero values, showing that they are not influenced by streamers. The values in this map, $W_b$, will be used for weighting later. This allows the definition of two regions in the observation, shown in figure \ref{refine}d. The red regions are strongly influenced by streamers, the green regions are not influenced by streamers. Each local region defined by a Gaussian function in longitude-latitude space maps to a set of points in the position-angle, time space of the observation. For example, the Gaussian function shown in figure \ref{refine}b maps to the points shown in figure \ref{refine}e. Furthermore, these sets of points can be separated into a subset of streamer and coronal hole points, shown as red and green respectively.

The ratio of the observed and reconstructed brightnesses is calculated, then a weighted mean of this ratio is calculated separately for the streamer and coronal hole points, with weightings given by $W_b$. For each Gaussian region of the density map, these two ratios are recorded. The global streamer and coronal hole corrections to be applied to the density map are calculated as the weighted sum of the streamer and coronal hole ratios from all the overlapping Gaussian regions, with the weights given by $W_{\rho}$. This results in two correction maps, one for streamers and one for coronal holes. The final correction map is created as follows. The mask of figure \ref{refine}b is smoothed over longitude-latitude space so that the streamer-coronal hole boundary is continuous. This smoothed mask gives a weighting at each point, $W$, that is used to calculate a weighted mean between the streamer and coronal hole correction factors. Thus densities in regions of the mask of figure \ref{refine}b which are far from streamers will have a correction applied that is given only by the coronal hole regions of the observation. Densities in regions near the streamer-coronal hole boundaries will have a correction which is a mean of the streamer/coronal hole correction. 

Once the correction is applied, the process is repeated, with a new streamer mask applied to the corrected density. With each iteration, the mean ratio of observed to reconstructed brightness becomes closer to unity, and the variance of the ratio becomes smaller. Convergence is reached when the changes in mean ratio and variance from one iteration to the next become small.

\section{Finding the optimal narrowing of streamers}
\label{app2}
\begin{figure}[]
\begin{center}
\includegraphics[width=8.0cm]{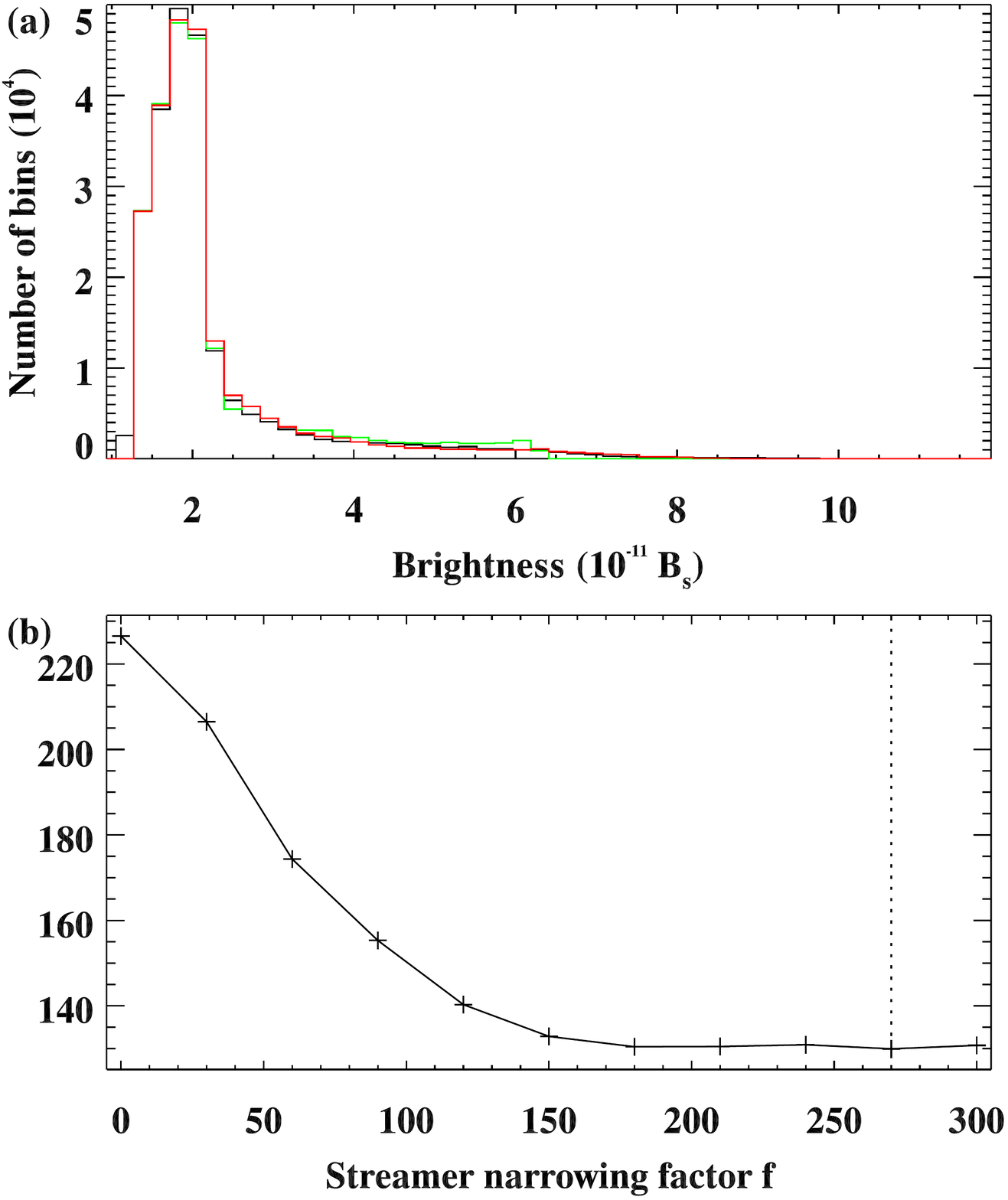}
\end{center}
\caption{(a) The observed brightness distribution is shown as a black line, and distributions gained from density reconstructions for values of $f$ of 0 (green) and 270 (red). (b) The weighted mean difference, $D$, between distributions as a function of the streamer narrowing factor, $f$. The minimum value of $D$ at $f=270$ is shown by the vertical dotted line.}
\label{optimise}
\end{figure}

Eleven values of the narrowing factor $f$ between 0 (no narrowing) and 300 are applied to the density, the refining method applied, and the resulting density and reconstructed brightness recorded for each value. A final step is to decide on the optimal narrowing factor. The black line of figure \ref{optimise}a shows the distribution of observed brightness. The green line shows the distribution of brightness given by the non-narrowed density ($f=0$). A weighted mean difference $D$ between the observed and reconstructed brightness distributions is given by
\begin{equation}
\label{eqb1}
D = \frac{\sum_i B_i |n_i - m_i|}{\sum_i B_i} , 
\end{equation}
where $B_i$ is the brightness at bin $i$, and $n_i$ ($m_i$) is the number of observed (reconstructed) values within this bin. This value is calculated for each streamer narrowing factor $f$. Figure \ref{optimise}b shows $D$ as a function of the narrowing factor $f$. The optimal value is given by the minimum of $D$, in this case $f=270$. The brightness distribution corresponding to $f=270$ is shown as the red line in figure \ref{optimise}a.

The use of equation \ref{eqb1} as a measure of goodness of fit requires some justification. The standard approach would be the mean of a point-by-point absolute difference, possibly weighted by the measurement errors. This would better fit the brightness of large, broad regions such as coronal holes and large, smooth streamers. However, this will be at the expense of finer-scale bright structures (such as narrow streamers), simply because the number of points occupied by these narrow structures is small. Thus the standard approach would give a small reduction in residuals over a large number of points, but a possibly large residual over a small number of points. This is one reason why equation \ref{eqb1} is used, which compares the brightness distribution rather than the spatial distribution, and sets a greater weighting to the small number of points which have a high brightness. Furthermore, equation \ref{eqb1} is forgiving in the sense that it is not important if there is a small difference in the position of a reconstructed brightness structure compared to the true corona (i.e. the spatial distribution is not considered). Note that the main spatial distribution has already been found using tomography, and that the streamer narrowing and density refinement steps are only adjustments to this initial model.
 
\end{document}